\documentclass[reprint,amsmath, twocolumn, amssymb, aps, notitlepage, superscriptaddress, pra]{revtex4-2}

\usepackage{amsmath,amsfonts}    % need for subequations
\usepackage{graphicx}   % need for figures
\usepackage{verbatim}   % useful for program listings
\usepackage{color,xcolor,soul}      % use if color is used in text
\usepackage{subfigure}  % use for side-by-side figures
\usepackage[colorlinks=true, citecolor=blue, urlcolor=blue]{hyperref}
\usepackage{textcomp}
\usepackage{bm}
\usepackage{bbm}
\usepackage{braket}
\usepackage[normalem]{ulem}
\usepackage{natbib}
\sethlcolor{yellow}

\definecolor{gr}{RGB}{225,225,225}

\renewcommand{\selectlanguage}[1]{}%to avoid language select error for a book entry

\begin{document}

% \title{Generating Entanglement and Squeezing in Chiral Bloch Domain Walls} 
% \title{Macroscopic stable entanglement in Chiral Domain Walls}
\title{Macroscopic entanglement between localized domain walls inside a cavity}

\author{Rahul Gupta}
\email[]{rahul.quantumfield@iitb.ac.in}
\affiliation{Department of Physics, Indian Institute of Technology Bombay, Powai, Mumbai 400076, India}

\author{H.Y. Yuan}
% \email[]{h.yuan1@uu.nl}
\affiliation{Institute for Advanced Study in Physics, Zhejiang University, 310027 Hangzhou, China}

\author{Himadri Shekhar Dhar}
% \email[]{himadri.dhar@iitb.ac.in}
\affiliation{Department of Physics, Indian Institute of Technology Bombay, Powai, Mumbai 400076, India}
\affiliation{Centre of Excellence in Quantum Information, Computation, Science and Technology, Indian Institute of Technology Bombay, Mumbai 400076, India}

\date{\today}

\begin{abstract}
We present a scheme for generating stable and tunable entanglement between two localized Bloch domain walls in nanomagnetic strips kept inside a chiral optical cavity. The entanglement is mediated by the effective optomechanical interaction between the cavity photons and the two macroscopic, collective modes of the pinned domain walls. By control of the pinning potential and optical driving frequency, the robust, steady-state entanglement between the two macroscopic domain walls can survive beyond the typical millikelvin temperature range.
\end{abstract}

% Further variation allows for enhancement with dispersion and even at finite temperatures. 
%%
% The system possesses many stable dissipative phases allowing for highly robust and tunable entanglement. We further show schemes based on parameter variation and choice of materials that can enhance the entanglement and its temperature cutoff. 
%%
% Collective dynamics of topological magnetic textures can be thought of as a massive particle moving in a magnetic pinning potential. We demonstrate that inside a cavity resonator this effective mechanical system can feel the electromagnetic radiation pressure from cavity photons through the magneto-optical inverse Faraday and Cotton-Mouton effects. We estimate values for the effective parameters of the optomechanical coupling for two spin textures: a Bloch domain wall and a chiral magnetic soliton lattice. The soliton lattice has magnetic chirality, so that in circularly polarized light it behaves like a chiral particle with the sign of the optomechanical coupling determined by the helicity of the light and chirality of the lattice. Most interestingly, we find a level attraction regime for the soliton lattice, which is tunable through an applied magnetic field.

% \pacs{}% insert suggested PACS numbers in braces on next line

\maketitle 

\section{Introduction\label{sec:intro}}
%\noindent\textbf{Introduction}\label{sec:intro}.--

The field of quantum magnonics~\cite{Bunkov2020,Yuan2022} has seen rapid growth, driven by the quest to harness the coherent properties of quantized collective spin waves in magnetic materials or {magnons} for emerging quantum technologies and solid-state device applications. 
For instance, hybrid quantum systems {\cite{LachanceQuirion2017,LachanceQuirion2019,Awschalom2021,Zhang2023,Yang2024}} that combine magnons with other quantum systems such as photonic cavities or superconducting qubits~{\cite{Xu2023,Xu2024,LachanceQuirion2020}} 
provide a platform to perform a variety of quantum operations~\cite{Flebus2023}, ranging from quantum computing~\cite{Andrianov2014} and storage~\cite{Simon2007} to quantum transduction~\cite{Pan2024}.

While much of the current focus in quantum magnonics has been on propagating magnons~\cite{Chumak2015}, localized magnetic spin textures such as chiral domain walls~\cite{Emori2013,Li2019}, skyrmions~{\cite{Finocchio2016,Petrovi2025}}, and spin vortices~\cite{Bogdanov1994} can offer additional advantages for quantum computing and related applications. These textures can be spatially pinned~\cite{Krumme1977-xo,Hanneken2016-lm,Holl2020} and arranged~\cite{Haller2022}, readily observed~\cite{Finco2021,Yu2021}, and also transported in magnetic racetrack circuits~{\cite{Parkin2015-rj,Zou2024}}. Furthermore, they often exhibit rich topological properties~\cite{Nagaosa2013-bh,Zhou2024}, making them attractive for both classical~\cite{Ababei2021,Lee2023,Park2024} and quantum~{\cite{Zou2023,Psaroudaki2023,Qu2025}} information processing.
% tasks.

An essential task in implementing key quantum protocols is to generate and control entanglement, which has been extensively explored in propagating magnon modes~\cite{Li2019a,Zhang2019,Luo2021,Yu2020,Yuan2020,Liu2023}. 
Most of these schemes rely on strong nonlinearity to generate entanglement either using Kerr effect~\cite{Zhang2019}, squeezed light~\cite{Yu2020} or optomechanical coupling in nonstationary magnon states~\cite{Luo2021}.
However, generating stable, steady-state entanglement in localized spin textures remains largely unexplored.  
%%%
Recent experimental progress has significantly improved the ability to manipulate spin textures such as a Bloch domain wall (DW), especially in uniaxial magnetic materials~\cite{Karna2021-es}.
These spin textures are already widely used in classical spintronic and nanoelectronic applications~\cite{Catalan2012,Yang2021-dy} and can operate at much lower energies compared to their electronic counterparts. Importantly, in the quantum regime, these DW behave as a macroscopic quasiparticle~\cite{Proskurin2018}, and extending their control into the quantum domain, especially by inducing entanglement and squeezing simply through external driving, opens up exciting new opportunities for quantum information processing, computing, and memory applications.

%%%%%%\comm{The line below should go in Discussion}
%\remove{Further, these systems operate at very low energy compared to their electronic counterparts, which is very desirable considering the rapidly increasing energy requirements for artificial intelligence \cite{Bourzac2024} and dealing with big data energy crisis.}
%%%%%%

In this work, we propose a compelling scheme to resolve the outstanding problem of  
entanglement in localized spin textures, by generating highly tunable and robust macroscopic entanglement between pinned Bloch domain walls in magnetic strips placed inside a chiral optical cavity~\cite{Voronin2022}. 
The photons inside the cavity create a radiation pressure on the two DWs through the inverse Faraday effect and generate an effective optomechanical interaction between the cavity mode and the {macroscopic DW} modes~\cite{Proskurin2018}.
%%%
To achieve stable entanglement, we exploit the theoretical machinations of cavity optomechanics~\cite{Aspelmeyer2014} to explore the dissipative phases of the system and identify phases with stable steady-state solutions, by tuning the pinning potential and the optical driving frequency.
%%
% The aim is to identify phases with stable steady-state solutions in dispersive regimes, with the magnon modes having either identical or different transition frequencies. 
%%
We observe that the {domain walls} are strongly entangled to the photonic mode close to the transition boundaries between two stable dissipative phases. The entanglement is robust across a broad range of optical power and can be controlled by the external field. For example, when {DW} frequencies are different, the driving can be varied to entangle either one of the {domain walls} with the photonic mode.
% transfer the entanglement between the photon mode and any one of the individual magnon modes. 
On the other hand, for identical {DW} frequencies, we can use the driving to continuously reduce the entanglement between the domain wall and photon modes in lieu of building strong entanglement between the two macroscopic, Bloch domain walls. 
%%
% On the other hand, for nonindentical magnon frequencies, the detuning can control and transfer the entanglement between the photon mode and the individual magnon modes.
%%
% For identical frequencies,  For instance, by changing the effective detuning, the entanglement between the magnon and photon can be diminished in lieu of strong entanglement between the two magnon modes. 
%%%%
Remarkably, the entanglement here is generated and controlled through the driven-dissipative properties of an optomechanical system rather than by introducing strong external nonlinearities in the dynamics.

The paper is arranged in the following way. In Sec.~\ref{sec:setup}, the setup and theoretical model for the interaction of DWs with a cavity field is proposed. Section~\ref{sec:snp} discusses distinct dissipative phases and their entanglement properties. This is followed by Sec.~\ref{sec:entangle}, where the origin of macroscopic DW entanglement and its thermal stability are presented. We conclude with a discussion and outlook in Sec.~\ref{sec:con}.

\begin{figure}[t]
    \includegraphics[width=\columnwidth]{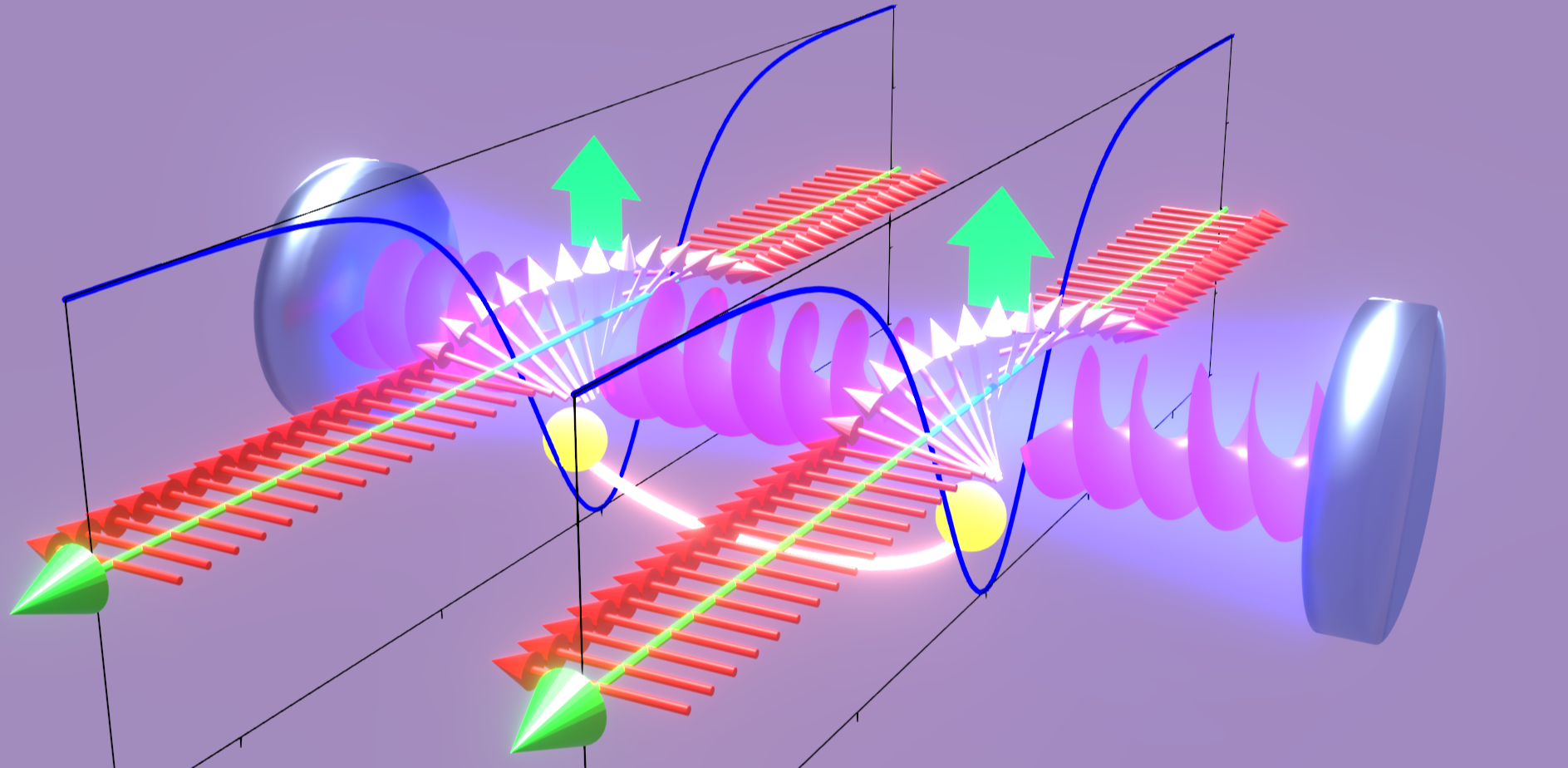}
    \includegraphics[width=\columnwidth]{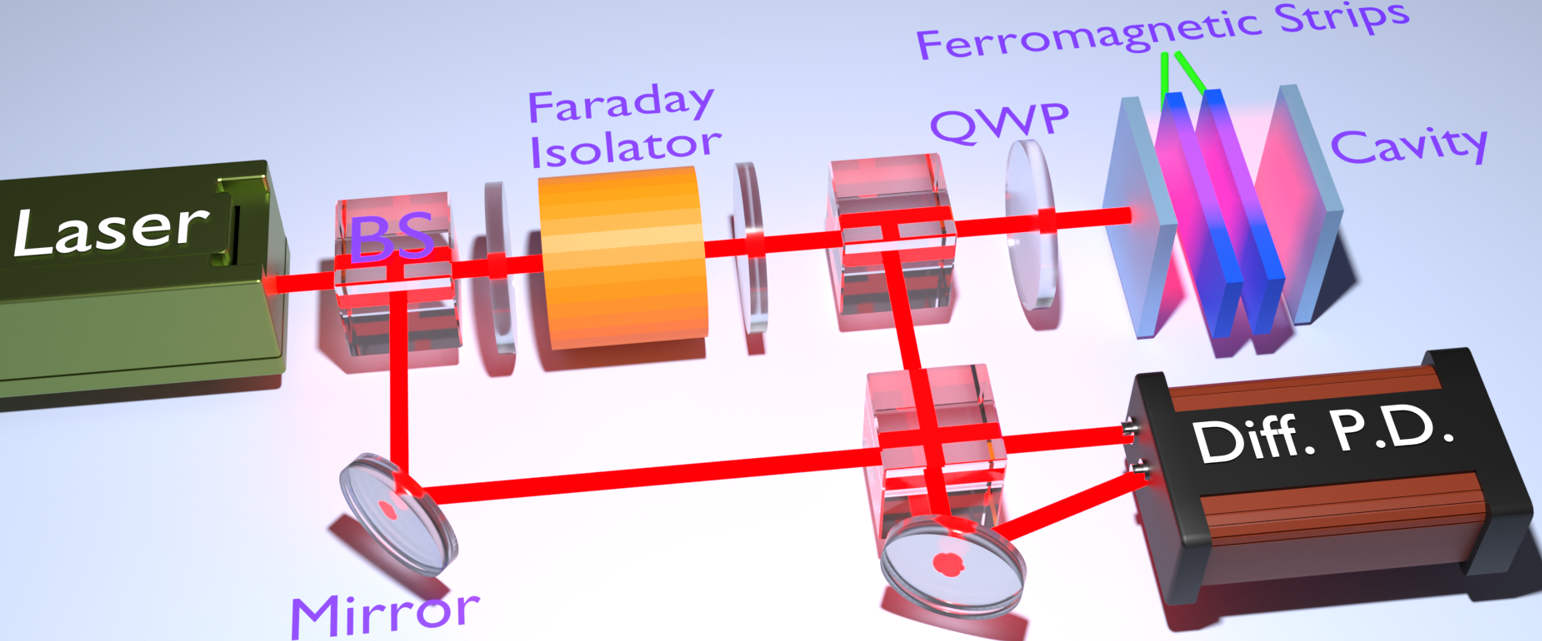}
    \caption{Experimental setup of domain walls inside a chiral cavity. The illustration shows linearly polarized light generated by a laser, which is passed through a Faraday isolator and a quarter-wave plate (QWP) to excite a circularly polarized optical mode. A right-circularly-polarized standing wave is created inside the chiral cavity, which then couples to the two ferromagnetic strips. The Bloch walls in the strips are pinned where the cavity field is present. The reflected optical mode is then collected at one port of a differential photodiode (Diff. PD) and the reference laser light is collected at the other port to obtain the spectrum via homodyne detection. The output and input are separated and merged with the use of a beam splitter (BS).}
    \label{fig:setup_model}
\end{figure}

\section{Setup and theoretical model\label{sec:setup}}

The primary objective of our work is to generate stable entanglement between two localized Bloch domain walls in spatially separated magnetic strips placed inside a single mode, chiral cavity. As such the setup of our protocol involves two key components viz., the domain wall and the chiral cavity.

Magnetic spin textures such as a domain wall (DW) occur naturally in a range of magnetic materials~\cite{Sherwood1959,Bobeck1967,malozemoff1979} and have rich topological structure that depend on the bulk properties and can majorly be classified in two types: Bloch walls and Neel walls~\cite{Thiaville2018}. These DWs can be localized spatially or pinned using various mechanisms \cite{Schafer1991,Kashuba1993,Jourdan2007,Koyama2011}.
Secondly, chiral cavities~\cite{Hbener2020} are designed to excite a single polarization state such as left or right circularly polarized light. Such single-handed cavities have recently been demonstrated~\cite{Voronin2022} and used for generating strong coupling with chiral molecules~\cite{Rosario2023}. An illustration of the setup involving Bloch DWs in ferromagnetic strips inside an optical cavity is shown in Fig.~\ref{fig:setup_model}.
%%
% Such cavities have recently been demonstrated \cite{Voronin2022} and have been used for generating strong coupling with chiral molecules \cite{Rosario2023}. 
% \hl{After passing through chiral cavity, the transmitted light from the input port is again passed to QWP, making it linearly polarized. This reflected beam is then diverted from a beam splitter (BS) and collected into a photodiode. The time-dependent intensity of the output photon number can thus be measured via a spectrum analyzer or digital oscilloscope.}
%%%%
The magnetic strips are placed in such a way that the polarized light inside the cavity passes through the center of the  pinned Bloch DWs. 
% The radiation pressure forces the pinned magnon modes of the DW to oscillate due to the inverse Faraday effect. 
%%
{The magneto-optical interaction between the cavity and the domain wall arises due to radiation pressure, as the circularly polarized light in the cavity 
generates an effective magnetic field in the medium, which can then couple to the spins inside the DW. 
This is known as the inverse Faraday effect~\cite{Pitaevskii1961}. 
While spin-orbit coupling mediates such effect in electronic systems~\cite{Pershan1966}, the interaction between the DW and cavity modes can be better understood in terms of how the electric permittivity tensor is modulated by spin oscillations.}
For a strong pinning potential, the DW mode behaves like a collective macroscopic particle and can be canonically quantized to resemble a quantum harmonic oscillator~\cite{Proskurin2018}.
% \begin{tabular}{||c | c c c||} 
%  \hline
%  Phase & k=2 & k=1 & k=0 \\
%  \hline
%  1 & 0 & 0 & 0 \\ 
%  \hline
%  2 & 0 & 0 & 1 \\
%  \hline
%  3 & 0 & 1 & 0 \\
%  \hline
%  4 & 0 & 1 & 1 \\
%  \hline
%  5 & 1 & 0 & 0 \\
%  \hline
%  6 & 1 & 0 & 1 \\
%  \hline
%  7 & 1 & 1 & 0 \\
%  \hline
% \end{tabular}
%The corresponding physical picture showing the mechanism is sketched in Fig.~\ref{fig:setup_model}.

\begin{figure*}[ht]
\centering
    \includegraphics[width=2\columnwidth]{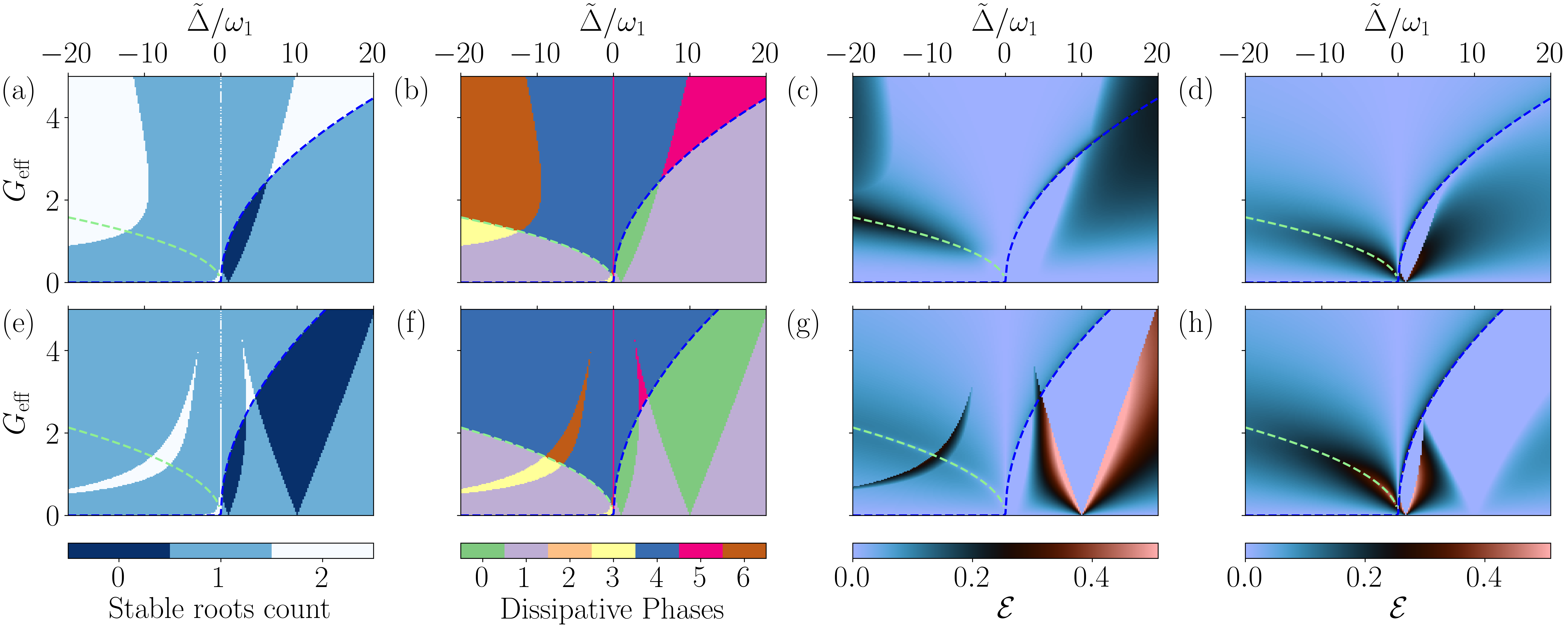}
    \caption{Dissipative phases and two-mode entanglement
    % . The figure shows the different phases and entanglement 
    of the system as a function of the detuning $\tilde{\Delta}$ and {$G_{\rm{eff}}=G |g_1|/\omega_1$} for $\omega_2=\omega_1$ (a-d) and $\omega_2=10~\omega_1$ (e-h).
    % : Detuning $\tilde{\Delta}$ is varied with driving strength $G$ for $\omega_2=\omega_1$ in (a-d) and $\omega_2=10\omega_1$ in (e-h).
    The plots (a) and (e) show the number of stable solutions, with the distinct dissipative phases {$p\in\{0,1,...,6\}$} shown in (b) and (f). The entanglement between the two DWs $\mathcal{E}_{1|2}$ and between the photon and DW modes $\mathcal{E}_{a|1(2)}$ are shown in plots (c) and (d). Finally, the plots (g) and (h) show the entanglement between the photon and DW modes $\mathcal{E}_{a|1}$ and $\mathcal{E}_{a|2}$ for $\omega_2=10~\omega_1$. 
% , (c,d) shows the bipartite DW-DW and photon-DW entanglement $\mathcal{E}_{1|2}$ and $\mathcal{E}_{a|1(2)}$ respectively for $\omega_1=\omega_2$, while (g,h) shows photon-DW entanglement $\mathcal{E}_{a|2}$ and $\mathcal{E}_{a|1}$ respectively for $\omega_2=10\omega_1$. 
The blue and green dashed lines highlight two interesting regions of phase transition. The former indicates the jump from a region with two to three real solutions, while the latter exhibits a line of transcritical bifurcation where  $\mathcal{E}_{1|2}$ in plot (c) is optimal. The parameters taken are $\kappa_a=2$~MHz, $\kappa_{1(2)}=1$~MHz, $\omega_1=1$~GHz, $g_{1(2)}=1$~MHz and temperature $T=2$~mK.}
    \label{fig:stability}
\end{figure*}

Under this quantized picture, the Hamiltonian of the system, neglecting the dissipation effects, can be written as $\mathcal{H} = \mathcal{H}_0 + \mathcal{H}_{\rm int} + \mathcal{H}_{L}$, where the terms represent the non-interacting, interacting and the driving, respectively. For $N$ ferromagnetic strips inside a single-mode chiral cavity, with frequencies $\omega_j$ and $\omega_a$, respectively, the non-interacting Hamiltonian is given by ($\hbar = 1)$,
\begin{equation}
\mathcal{H}_0 = \omega_a a^\dag a + \sum_{j=1}^{N}\omega_j b_j^\dag b_j,
\label{non_interact}
\end{equation}
where $a$ and $b_{1}$,$b_{2},...,b_{N}$ are the bosonic operators for the cavity and the $N$ magnon modes. Each DW is pinned to a frequency $\omega_j=\sqrt{2K^j_{\rm pin}K_\perp l/\hbar^2 \lambda_{\rm DW}}$ that can be individually controlled by varying the pinning field. Here, $K^j_{\rm pin}$ is the anisotropic pinning field strength, $K_{\perp}$ is the out of plane anisotropy parameter of the magnetic insulator with unit cell dimension $l$, and $\lambda_{\rm DW}$ is the effective DW length.
%%
%%
% In this work, we will use the quantized description only.
% We take the cavity mode as $a$ and call a total of $N$ DW modes for $N$ ferromagnetic strips as $b_{1}$,$b_{2},...,b_{N}$. We write the Hamiltonian of the full system, neglecting the dissipation effects, in three components as $\mathcal{H} = \mathcal{H}_0 + \mathcal{H}_{\rm int} + \mathcal{H}_{L}$, where $\mathcal{H}_0/\hbar = \omega_a a^\dag a + \sum_{j=1}^{N}\omega_j b_j^\dag b_j$ is the non-interacting Hamiltonian containing a single cavity mode at resonant frequency $\omega_a$ and $N$ DW modes, each pinned to a frequency $\omega_j=\sqrt{2K^j_{\rm pin}K_\perp a/\hbar^2 \lambda_{\rm DW}}$ which is determined by (as shown in \cite{Proskurin2018} for single DW) the strength of the magnetic anisotropic pinning strength $K^j_{\rm pin}$ and out of plane anisotropy $K_{\perp}$ of the magnetic insulator material with unit cell dimension $a$ and effective domain wall length scale $\lambda_{\rm DW}$.
% \comm{Comm: The above statement is not clear without explaining what the domain wall is -- a reader will not be able to visualise/imagine what these various quantities are.}
%%
%%
The interaction \cite{Proskurin2018,Aspelmeyer2014} and driving Hamiltonian $\mathcal{H}_1=\mathcal{H}_{\rm int}+\mathcal{H}_{L}$ is given by 
\begin{equation}
\mathcal{H}_1=\sum_{j=1}^{N}g_j a^\dag a(b_j^\dag+ b_j) + i(a^\dag \xi e^{-i\omega_{L} t} - a \xi^* e^{i\omega_{L} t}),
\label{eq:H_drive}
\end{equation}
% $\mathcal{H}_{\rm int}/\hbar=\sum_{j=1}^{N}g_j a^\dag a(b_j^\dag+ b_j)$, 
where $g_j$ is the coupling between a photon and the $j^{th}$ magnon mode arising due to inverse Faraday effect and $\xi$ is the strength of the driving field with frequency $\omega_{L}$.
The magneto-optic coupling term is given by $g_{j}=-c\phi_F\sqrt{\epsilon}S^j_{\rm eff}/2$, and creates an effective optomechanical interaction between the cavity and magnon modes. 
% where
%%
% which contains the magneto-optic coupling-induced single photon to DW interaction via inverse Faraday effect. This is very similar to optomechanical interaction, however, the single photon-DW coupling strength $g_{j}=-c\phi_F\sqrt{\epsilon}S^j_{\rm eff}/2,~\forall j\in\{1,2,...,N\}$, 
% is independent of $\omega_j$ and directly depends on the Faraday rotation 
{This is independent of the cavity frequency $\omega_a$, unlike the usual optomechanical coupling where $g_j\propto\omega_a$ \cite{Aspelmeyer2014},} and {is} dependent on the Faraday rotation $\phi_F$, which is the amount of rotation in polarization per unit length as light passes through a material of dielectric permittivity $\epsilon$. 
% , and is fixed for a given material. $\epsilon$ is the dielectric permittivity of the material.
$S^j_{\rm eff}$ is a dimensionless geometrical factor that depends on the scattering cross-section, cavity volume and zero point fluctuation. See Appendices~\ref{app:quantization} and \ref{app:parameters} for more details. The higher-order terms and non-linear Kerr effect have been neglected. 
%%
% The dimensionless geometrical factor $S^j_{\rm eff}=x^j_{\rm zpf}A^j_{\perp}/V_c$ depends
% % that takes into account of the available 
% on the scattering cross-section $A^j_{\perp}$, 
% % where the cavity field penetrates it, 
% the cavity volume $V_c$ and the zero point fluctuation $x^j_{\rm zpf}=\sqrt{\hbar/2m_j\omega_j}$ of the macroscopic domain wall, with center of mass $m_j=\rho V_j$, where $\rho$ is the density of the material and $V_j$ is the effective cross-sectional volume of the DW. We note that higher-order terms and non-linear Kerr effect have been neglected. 
% %%%
% The final term in the Hamiltonian corresponds to the driving,
% $\mathcal{H}_{L}/\hbar= i(a^\dag \xi e^{-i\omega_{L} t} - a \xi^* e^{i\omega_{L} t})$ contains the input pump laser field with strength $\xi$ at frequency $\omega_{L}$. The pumping rate is related to the input power $P_{\rm in}$ via the relation $|\xi|=\sqrt{2P_{\rm in}\kappa_a/\hbar\omega_L}$, where $\kappa_a$ is the cavity dissipation rate.
% Similarly one can account for Gilbert damping and cavity losses via a non-hermitian component
% \begin{align}
%     \mathcal{H_{\rm NH}}/\hbar &= i\kappa_a a^\dag a + i\sum_{j} \kappa_j b_j^\dag b_j\\
%     \kappa_j&=\frac{\alpha_G \omega_j}{\sqrt{K^j_{\rm pin}a/2K_{\perp}\lambda_{\rm DW}}}
% \end{align}
% Here, $\alpha_G$ is the Gilbert damping rate of the material.
In the frame moving with the driving field i.e., under the unitary transformation $\mathcal{H'} = \mathcal{U}\mathcal{H}\mathcal{U}^{\dagger} - i\hbar \mathcal{U} \frac{\partial \mathcal{U}^{\dagger}}{\partial t}$, 
where $\mathcal{U}=e^{i \omega_{L} a^\dag a t}$, the Hamiltonian is time-independent
% The explicit time dependence in this Hamiltonian can be removed by shifting to the frame oscillating with the laser by applying unitary transformation with $\mathcal{U}=e^{i\omega_{L} a^{\dagger}a t}$, such that
\begin{align}
    % \mathcal{H'} &= U\mathcal{H}U^{\dagger} - i\hbar U \frac{\partial U^{\dagger}}{\partial t},\nonumber\\
    \mathcal{H'} &= -\Delta_{a}a^{\dagger}a + \sum_{j=1}^{N}\left[ \omega_j b_j^\dag b_j + g_j(b_j^\dag+ b_j)a^\dag a\right]\nonumber\\
    &+ i\xi~ (a^\dag - a),%~;~\Delta_{a} = \omega_{L}-\omega_{a}
\label{eq:H'}
\end{align}
where $\Delta_{a} = \omega_{L}-\omega_{a}$, is the cavity detuning.
\section{Stable Entangled Phases \label{sec:snp}}

Since our goal is to generate stable steady-state entanglement, it is important to first 
% perform a stability analysis and 
identify the driven dissipative phases of the system with stable solutions~\cite{Bibak2023}.  
% that can be achieved within the experimentally feasible parameter regime.
%%
To this effect, the Hamiltonian in Eq.~\eqref{eq:H'} is linearized around the steady state photon number $\bar{n}_a$. 
In the limit of strong driving, 
% the Hamiltonian in Eq.~\eqref{eq:H'}, %, the detailed analysis of various dissipative phases for a single mechanical mode ($N=1$) was carried in \cite{Bibak2023}
% in the limit of strong driving, the dissipative phases 
% can be studied by linearizing the system around the steady state photon number 
$\bar{n}_a$ is obtained as the roots of the cubic equation 
% for the steady-state photon number, 
$\bar{n}_a=\vert\xi\vert^2/\left[\left(\Delta_a +\Omega\bar{n}_\alpha\right)^2 + \kappa_a^2\right]$, where $\Omega=\sum_{j=1}^{2}2g^2_j\omega_j/(\omega^2_j+\kappa^2_j)$, and $\kappa_a$ and $\kappa_j$ are the dissipation rates in cavity and $j^{\rm{th}}$ DW mode, respectively.
%%
% For the $k^{th}$ root, 
% % Indexing the roots using $k$, 
% where $k\in\{1,2,3\}$, 
The steady states of the different operators are written as $\langle a\rangle = \alpha^{(k)}$, $\langle a^\dag a\rangle=\bar{n}_a^{(k)}$, and $\langle b_j\rangle=\beta_j^{(k)}$, where $k\in\{0,1,2\}$ indicate the roots (see Appendix~\ref{sec:appA} for the derivation).
Let $G^2\equiv\bar{n}^{(k=0)}_a$ denote the real root 
% of cubic equation 
common to all parameter regimes, with
$G={\vert\xi\vert}/{\sqrt{\tilde{\Delta}^2 + \kappa_a^2}}$, where 
% $\tilde{\Delta}\equiv\tilde{\Delta}^{(k=0)}=\Delta - \Omega G^2$.
$\tilde{\Delta}=\Delta_a + \Omega G^2$.
The advantage of this representation is that 
% the steady state photon number 
$\bar{n}_a^{(k)}$ corresponding to the other roots can be parametrized in terms of $G$ and $\tilde{\Delta}$, such that for $k\in\{1,2\}$
% using the sum and product of cubic roots such that for $k\in\{1,2\}$,
%%%
% one can directly parameterize the remaining roots of $\bar{n}_a^{(k)}$ with $\tilde{\Delta}$ and $G$ by a simple quadratic equation
\begin{equation}
    \bar{n}_a^{(k)}=\frac{G^2}{2} - \frac{\tilde{\Delta}}{\Omega} + (-1)^{k-1}\sqrt{\left(\frac{G^2}{2} - \frac{\tilde{\Delta}}{\Omega}\right)^2 - \frac{\kappa_a^2 + \tilde{\Delta}^2}{\Omega^2}}.
    \label{roots}
\end{equation}
Since complex roots appear in conjugates, the system can have either one or three real roots. For the latter,
% existence of 3 real roots  $\bar{n}_a^{(k)}$, 
the parameters must satisfy the positive discriminant condition in Eq.~\eqref{roots}, 
% which gives us
$G^2\geq 2{\Omega}^{-1}({\tilde{\Delta}}+\sqrt{{\kappa_a^2 + \tilde{\Delta}^2}}).$
The blue-dashed line separating the two regimes is shown in Fig.~\ref{fig:stability}.
% These two roots for $k\in\{1,2\}$ emerge only when
% \begin{equation}
% \frac{G^2}{2}\geq\frac{\tilde{\Delta}}{\Omega}+\sqrt{\frac{\kappa_a^2 + \tilde{\Delta}^2}{\Omega^2}}.
% \label{eq:limits}
% \end{equation}
%%%
% The detailed derivation of these states is carried out in App.~\ref{sec:appA}, 
%%%

Let us now focus on the case of two ferromagnetic strips ($N=2$) inside the cavity. 
The DW in each strip is pinned with different potentials leading to two magnon modes with frequencies $\omega_1$ and $\omega_2$. The effective Hamiltonian linearized around $\bar{n}^{(k)}_a$ is given by 
%We have performed this analysis to obtain a linearized description of Hamiltonian for all three steady-state values in .
\begin{align}
H_{\rm eff}&=-\tilde{\Delta}^{(k)}a^{\dag}a + \omega_1 b^\dag_1 b_1 + \omega_2 b^\dag_2 b_2 +\left(\alpha^{(k)} a^{\dag}+ \alpha^{(k)*} a\right) \nonumber\\
&\times \left[g_1(b_1 + b^\dag_1) +g_2(b_2 + b^\dag_2)\right],
\label{eq:H_collective}
\end{align}
where $\tilde{\Delta}^{(k)}=\Delta_a-2\text{Re}(\beta_j^{(k)})$. 
To analyze the stability of the dissipative phases, we study the 
linearized 
quantum Langevin equations (QLE)~\cite{Aspelmeyer2014} for $u = \left[\{x_i,y_i\}\right]$ for $i\in\{a,1,2\}$, where $x_i= (\mathcal{O}_i+\mathcal{O}^\dag_i)/\sqrt{2}$ and $y_i= i(\mathcal{O}_i-\mathcal{O}^\dag_i)/\sqrt{2}$ are the quadrature operators for $\mathcal{O}_i \in [a,b_1,b_2]$. 
See Appendix~\ref{app:adia} for the equations.

%%
% around the $k^{th}$ steady state, $u^{(k)} = \left[\{x^{(k)}_i,y^{(k)}_i\}\right]$, where $x^{(k)}_i=(o^{(k)}_i+o^{\dag(k)}_i)/\sqrt{2}$, $y^{(k)}_i=i(o^{\dag(k)}_i-o^{(k)}_i)/\sqrt{2}$, $\eta = \left[\{\sqrt{2\kappa_i}~x^{\rm in}_i,\sqrt{2\kappa_i}~p^{\rm in}_i\}\right]$, where $i \in [a,1,2]$ and $o_i \in [a,b_1,b_2]$. 
For the $k^{th}$ root, the QLE is given by
$\dot{u}^{(k)}=A^{(k)}u^{(k)} + \eta$, where $\eta = \left[\{\sqrt{2\kappa_i}~x^{\rm in}_i,\sqrt{2\kappa_i}~p^{\rm in}_i\}\right]$ and
\begin{align}
A^{(k)}=
\begin{pmatrix}
-\kappa_a & -\tilde{\Delta}^{(k)} & -\tilde{g}_1y^{(k)}_\alpha & 0 & -\tilde{g}_2y^{(k)}_\alpha & 0\\
\tilde{\Delta}^{(k)} & -\kappa_a & \tilde{g}_1x^{(k)}_\alpha & 0 & \tilde{g}_2x^{(k)}_\alpha & 0\\
0 & 0 & -\kappa_1 & \omega_1 & 0 & 0\\
\tilde{g}_1x^{(k)}_\alpha & \tilde{g}_1y^{(k)}_\alpha & -\omega_1 & -\kappa_1 & 0 & 0\\
0 & 0 & 0 & 0 & -\kappa_2 & \omega_2\\
\tilde{g}_2x^{(k)}_\alpha & \tilde{g}_2y^{(k)}_\alpha & 0 & 0 & -\omega_2 & -\kappa_2
\end{pmatrix},\label{eq:drift_mat}%\nonumber
\end{align}
where $x(y)_\alpha^{(k)}=\langle x(y)_a^{(k)}\rangle$ and $\tilde{g}_{1(2)} = g_{1(2)} \sqrt{2}$.
% , for $j \in \{1,2\}$.
%%
% $u = \left[x_a,p_a,x_1,p_1,x_2,p_2\right]^T$, and 
% \begin{align}
% \eta &=\left[\sqrt{2\kappa_a}x^{\rm in}_a,\sqrt{2\kappa_a}p_a^{\rm in},\sqrt{2\kappa_1}x_1^{\rm in},\sqrt{2\kappa_1}p_1^{\rm in},\sqrt{2\kappa_2}x_2^{\rm in},\right.\nonumber\\
% &\left.\sqrt{2\kappa_2}p_2^{\rm in}\right]^T.\nonumber
% \end{align}
% $\eta = \left[\sqrt{2\kappa_a}x^{\rm in}_a, \sqrt{2\kappa_a}p_a^{\rm in}, \sqrt{2\kappa_1}x_1^{\rm in}, \sqrt{2\kappa_1}p_1^{\rm in},\sqrt{2\kappa_2}x_2^{\rm in},\sqrt{2\kappa_2}p_2^{\rm in}\right]^T$.
%%
% \hsd{Rahul: We need to describe what quadrature of fluctuations $u^{(k)}$ means?-done It seems it is not fluctuation of quadrature. Also check if we are keeping notation such as $G^2$ below.-to be think}
%
The stable phases of the system are 
% ility of the system dynamics is 
related to the convergence of the system to a steady state, which requires that $\rm max\{Re(\{\lambda_A\})\}\leq0$, where $\{\lambda_A\}$ are the eigenvalues of the matrix $A^{(k)}$.

%
% Now the condition for stability is $\rm max\{Re(\{\lambda_A\})\}\leq0$ where $\{\lambda_A\}$ is the set of eigenvalues of $A^{(k)}$ matrix. This ensures the convergence to steady state values in dynamics. 
% \hsd{Perhaps we can explain in words why this is the condition for stability. Also, why is there a root that is common in all parameter regimes.-done}
%%%

In the strong driving regime ($\bar{n}_a \gg 1$), the system is Gaussian and the  
entanglement between the three modes $a,b_1$ and $b_2$ 
% in the linearized Gaussian regime $G\ll\xi$, the entanglement 
can be quantified by the properties of the 
% calculating 
steady state covariance matrix $V^{(k)}$ of the system. 
% which can be extracted 
This can be obtained by solving the Lyapunov equation
\begin{align}
&A^{(k)}V^{(k)} + V^{(k)} A^{(k)T}=-D,
% &D=\left[\kappa_a(2n^{\rm th}_a +1),\kappa_a(2n^{\rm th}_a +1),\kappa_1(2n^{\rm th}_1 +1),\right.\nonumber\\
% &\left.\kappa_1(2n^{\rm th}_1 +1),\kappa_2(2n^{\rm th}_2 +1),\kappa_2(2n^{\rm th}_2 +1)\right]^T\\
% &n^{\rm th}_i=\frac{1}{e^{\hbar\omega_i/k_B T} - 1}~~;~~\forall i \in \{a,1,2\}
\label{eq:lyapu}
\end{align}
where $D=[\{\kappa_i(2n^{\rm th}_i +1),\kappa_i(2n^{\rm th}_i +1)\}]$ and $n^{\rm th}_i={1}/(e^{\hbar\omega_i/k_B T} - 1)$ for $i\in\{a,1,2\}.$ 
%%%%
% \begin{figure*}[ht]
% \centering
%     \includegraphics[width=2\columnwidth]{figures/phases_x_mixed.png}
%     % \includegraphics[width=2\columnwidth]{figures/phases.png}
%     % \includegraphics[width=2\columnwidth]{figures/spectra.png}
%     \caption{Identification of dissipative phases: Number of stable solutions (a,e) and dissipative phases (b,f) plotted with detuning $\tilde{\Delta}$ and driving rate $G$. The top panel (a-d) shows results for $\omega_2=\omega_1$, while the bottom panel shows the results for $\omega_2=10\omega_1$. (c,g) and (d,h) shows the output optical spectrum for the root $k=1$ and $k=2$ respectively. The dotted, dense green curve shows the eigenvalues across the spectral maxima.}
%     \label{fig:stability}
% \end{figure*}
%%%
To quantify entanglement between any two of the modes (say the two DW modes), corresponding to the stable root $k$, we eliminate the third mode (the photon mode) from the covariance matrix. The reduced 
% the two DW modes, for the stable roots $k$, we eliminate the photon mode from the $V^{(k)}$ matrix to get a 
$4\times 4$ covariance matrix for the two DW modes $\tilde{V}^{(k)}_{1|2}$ is then used to study entanglement.
The measure used is logarithmic negativity (LN), an entanglement monotone based on Simon's criterion~\cite{Simon2000}, given by 
% , from which we obtain the logarithmic negativity to quantify entanglement, which is obtained as 
$\mathcal{E}_{1|2}=\rm{max}\{0,\ln(2\mu_{-})\}$, where $\mu_-$ is the lowest eigenvalue of the symplectically transformed reduced covariance matrix.
% $\tilde{V}^{(k)}_{12}$ and is an entanglement monotone using Simon's criterion \cite{Simon2000}. 
Similarly, the entanglement $\mathcal{E}_{a|j}$ between the photon and DW mode $j$ is obtained by estimating LN for the reduced covariance matrix $\tilde{V}^{(k)}_{a|j}$. 
% and repeat the same process.

Figure~\ref{fig:stability} illustrates the stable phases of the driven-dissipative system and the entanglement between the different modes, respectively. 
%%
% \textcolor{red}{The phases are classified based on the stability $s_i$ of the $i^{\rm{th}}$ root, such that $s_i=0(1)$ for stable(unstable) root, and we assign a phase number $P=\sum_{i=0}^2 2^i s_i$, which is decimal equivalent of the stability bit string $\{s_2,s_1,s_0\}$.} 
%
%
These phases are for different values of {effective coupling $G_{\rm{eff}}=G |g_1|/\omega_1$} and the detuning $\tilde{\Delta}$, which can be controlled by the driving field strength and frequency, {and obtained by solving the mean-field QLE.}
Moreover, we consider the cases where the two DW magnon modes are either resonant ($\omega_2 = \omega_1$) or considerably detuned ($\omega_2 = 10\omega_1$), depending on the choice of the pinning field strength at the magnetic strips. 
%%
%%%
% The number of stable phases are shown in Figs.~\ref{fig:stability}(a,e), 
% and \ref{fig:stability}(e), 
% where a red line demarcates the boundary between phases with one and three real roots based on Eq.~\eqref{roots}. 
% whereas the different dissipative phases of the system are shown in Figs.~\ref{fig:stability}(b,f). 
% and \ref{fig:stability}(f).
% We plot the count of the total number of stable roots $\bar{n}^{(k)}_a$ in Fig.~\ref{fig:stability}(a,e) by tuning the driving via $G\propto|\xi|$ and the effective detuning $\tilde{\Delta}$ to broadly categorized the distinct dissipative phases as shown in Fig.~\ref{fig:stability}(b,f). 
% We find that one can control the mono and bistable regimes by simply controlling the ratios of the two resonant frequencies, which can be altered by varying pinning field strength at one of the strips. 
%%
The number of stable phases and
% are shown in Figs.~\ref{fig:stability}(a,e), 
% and \ref{fig:stability}(e), 
% where a red line demarcates the boundary between phases with one and three real roots, and
% based on Eq.~\eqref{roots}. 
% The 
the different dissipative phases are shown in {Figs.~\ref{fig:stability}(a,e) and (b,f), respectively, while
Figs.~\ref{fig:stability}(c) and (d,g,h) shows the DW-DW and photon-DW} entanglement in the different phases of the system, maximized over all stable roots.
%%%
{The different phases in Fig.~\ref{fig:stability}(b,f) are based on which of the three roots ($k \in \{0,1,2\}$) are stable. Using binary digits, we can identify whether the $k^{th}$ root is stable ($s_k = 1$) or not ($s_k = 0$). The dissipative phase is then given by the number $p = \sum_k 2^k s_k$. As there are no phases where all three roots are stable, so the phases range from 0 up to 6. Note that the mean photon number $\bar{n}_a^{(k)}$ is an order parameter for the dissipative phase transitions~\cite{Bibak2023}.}

\begin{figure}[t]
	\includegraphics[width=\columnwidth]{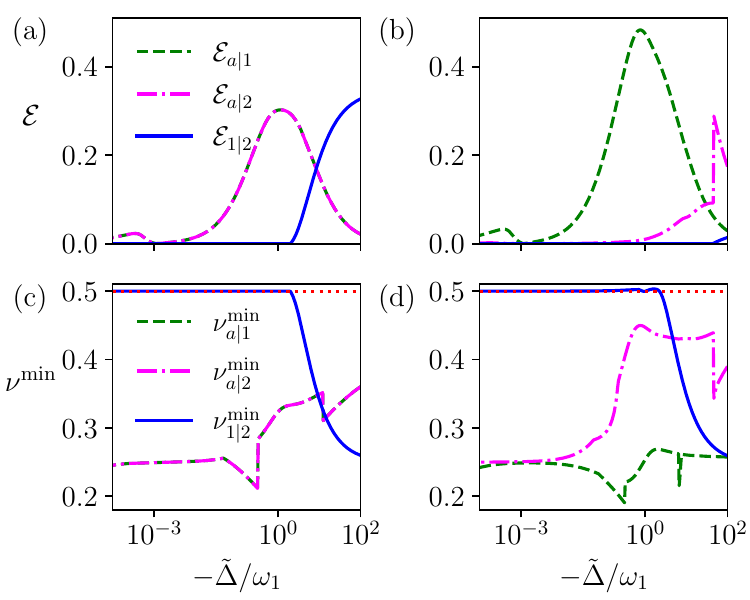}
	\caption{Entanglement switching and two-mode squeezing. The figure shows (top) bipartite entanglement $\mathcal{E}_{i|j}$ and (bottom) lowest eigenvalue $\nu^{\rm{min}}_{i|j}$ of the covariance matrix $\tilde{V}^{(k)}_{i|j}$, between modes $i$ and $j$ (where $i,j\in\{a,1,2\}$). The plots are for $\omega_1=\omega_2$ (a,c) and $\omega_2=10~\omega_1$ (b,d). 
    % Logarithmic negativity (a,b) as entanglement measure and squeezing (c,d) for photon-DW modes $\mathcal{E}_{a|1(2)}, \nu^{\rm{min}}_{a|1(2)}$ (green and pink dashed curves) and DW-DW modes $\mathcal{E}_{1|2},\nu^{\rm{min}}_{1|2}$ respectively. (a,c) are for $\omega_1=\omega_2$ while (b,d) are for $\omega_2=10\omega_1$. 
    All other parameters are same as in Fig.~\ref{fig:stability}.
    }
    \label{fig:ent_phases}
\end{figure}

Now, the key objective of the work is to demonstrate significant entanglement between the two DW magnon modes. In this regard, it is worth exploring regions where transcritical bifurcation occurs i.e., point where two of the roots exchange their stability (the stable root becomes unstable, while unstable root becomes stable), thus keeping the total number of stable roots fixed and marking a continuous dissipative phase transition. Let us consider the case $\bar{n}_a^{(0)}=\bar{n}_a^{(k)}$ for $k \in \{1,2\}$, where such a
%
% The important region that we are interested in exploring is when $\bar{n}_a^{(0)}=\bar{n}_a^{(1,2)}$, as here a transcritical bifurcation occurs such that two of the roots exchange their stability, the stable root becomes unstable, while unstable root becomes stable, keeping the total number of stable roots fixed, thus marking a continuous dissipative phase transition. This 
%
transition occurs at $G=\sqrt{-(\kappa_a^2 + \tilde{\Delta}^2)/2\tilde{\Delta}\Omega}$ for $\tilde{\Delta}\leq 0$. These bifurcations are highlighted in Fig.~\ref{fig:stability} by a green dashed line. The entanglement properties of the system along this line are remarkable. For $\omega_1=\omega_2$, the entanglement between photonic modes and the domain wall, $\mathcal{E}_{a|j}$ for $j \in \{1,2\}$, is maximum close to $\tilde{\Delta}=-\omega_{1}$, as shown in Figs.~\ref{fig:stability}(c,d). Note that the photon mode is equally entangled with both the DW modes. However, moving away to the left along the transition line ($\tilde{\Delta} < 0$), the photon-DW entanglement decreases and entanglement between the two DW modes $\mathcal{E}_{1|2}$ becomes significantly stronger. {This transition from photon-DW to entanglement between the two magnetic textures occurs at $-\tilde{\Delta}/\omega_1 = \kappa_a/\kappa_1$, and is captured in {Fig.~\ref{fig:ent_phases}(a)}.} 
%%%
Incidentally, for unequal pinning strengths $\omega_2=10\omega_1$, the entanglement between the two DWs is suppressed. Moreover, the photon mode entangles with only one of the DW modes. For instance in Fig.~\ref{fig:ent_phases}(b), the photon mode entangles with the first DW mode at {$\tilde{\Delta}=-\omega_1$} i.e., $\mathcal{E}_{a|1}$ is finite. Again by moving along the bifurcation line towards {$\tilde{\Delta}=-\omega_2$}, the entanglement can switch to the second DW mode and $\mathcal{E}_{a|2}$ is now significant. As such, by controlling the pinning strengths and the detuning, entanglement can be generated between the two DW modes or between the photon and a specific DW mode. Note that the tradeoff of entanglement is enforced by the monogamy principle~\cite{Terhal2004,Hiroshima2007}, which restricts the presence of significant entanglement simultaneously between all possible pairs of modes.

{A physical interpretation of the existence of entanglement along the bifurcation line can be provided by analyzing the correlations between the optical and DW modes i.e., the covariances $\langle x_i x_j\rangle$, $\langle x_i y_j\rangle$ and $\langle y_i y_j\rangle$ for $i \in \{a,1,2\}$.
In the vicinity of dissipative phase transition, correlations and decay rates tend to diverge~\cite{Horstmann2013}. Our numerical results~\footnote{All simulations and illustrations were done with the help of open source Python libraries~\cite{Lambert2024} and repositories for the software Blender~\cite{Blender2023,Mitzua_2023}, respectively.} 
show that correlation between the cavity and DW modes $\langle p_a x_1\rangle$ or the DW-DW modes $\langle x_1 x_2\rangle$ are often large and comparable to local mode fluctuations such $\langle p_a p_a\rangle$ or $\langle x_1 x_1\rangle$. This allows for build up large quadrature correlations and entanglement, as quantified by the logarithmic negativity. A qualitative discussion on the variation of correlation and local fluctuation near the bifurcation line is presented in Appendix~\ref{app:ent}.}

\section{Macroscopic entanglement\label{sec:entangle}}
% In this section, we qualitatively determine why large detuning regions of parameter space lead to DW-DW entanglement despite there is no direct interaction in Hamiltonian, and we explore what parameters can lead to its enhancement.

While entanglement between the photonic and DW modes arise from the direct light-matter coupling in the linearized Hamiltonian in Eq.~\eqref{eq:H_collective}, the generation of entanglement between the two macroscopic DW modes in the absence of any nonlinearity is not straightforward. 
In our work, this is achieved by using strong coupling to move to an dispersive regime in terms of the effective detuning $\tilde{\Delta}$ from the cavity modes. Mathematically this entails that  $\tilde{\Delta}\gg \omega_{1},\kappa_{a}$.
%
% Since the Hamiltonian in Eq.~\eqref{eq:H_collective} doesn't has any direct coupling term for the two DW modes, thus to achieve effective coupling one utilizes the strong coupling regime where we have a large effective detuning ($\tilde{\Delta}\gg \omega_{1},\kappa_{a}$) from the cavity resonance mode. 
In this limit, we adiabatically eliminate the cavity mode using the steady-state value of the $k^{th}$ root and the relation with the DW mode fluctuations
$a = {i\alpha\sum_j g_j(b_j + b^\dag_j)}/{(i\tilde{\Delta}^{(k)} -\kappa_a)}$. The 
% resulting adiabatic Hamiltonian 
resulting Hamiltonian (derived in Appendix~\ref{app:adia}) is given by 
\begin{align}
    H_{\rm ad} =\sum_j \omega^{\rm eff}_j b^\dag_j b_j + \sum^{j\neq i}_{i,j}G^{\rm eff}_{ij}\left(b_i+b^\dag_i\right)\left(b_j + b^\dag_j\right),\label{eq:H_ad}
    % H_{\rm ad}&=\sum_j \omega^{\rm eff}_j \delta b^\dag_j b_j + \sum^{j\neq k}_{j,k}G^{\rm eff}_{jk}\left(\delta b^\dag_j \delta b_k + \delta b_j \delta b^\dag_k\right)\nonumber\\
    % &+ \sum^{j\neq k}_{j,k}G^{\rm eff}_{jk}\left(\delta b^\dag_j \delta b^\dag_k + \delta b_j \delta b_k\right).\label{eq:H_ad}
\end{align}
where, the effective mode frequency is  $\omega^{\rm eff}_j=\omega_j + G^{\rm eff}_{jj}$, with $G^{\rm eff}_{ij}={\tilde{\Delta}^{(k)}\vert\alpha\vert^2 g_i g_j}/{(\tilde{\Delta}^{(k) 2} + \kappa_a^2)}$. 
% The derivation is given in Appendix~\ref{app:adia}.
% \begin{align}
%     \omega^{\rm eff}_j&=\omega_j + G^{\rm eff}_{j,j}~;~G^{\rm eff}_{j,k}=\frac{\tilde{\Delta}^{(m)}\vert\alpha\vert^2 g_j g_k}{\tilde{\Delta}^{(m) 2} + \kappa_a^2}
% \end{align}

The coupling between the two DW modes $b_i$ and $b_j$ are governed by  $G^{\rm eff}_{ij}$, which 
% for a fixed photon number 
increases with the effective detuning $\tilde{\Delta}^{(k)}$ before eventually saturating. Note that the interaction includes both energy conserving and non-conserving terms, with the latter giving rise to squeezing of the collective modes. This is closely connected to the macroscopic entanglement properties of the system. 
% {The onset of the continuous dissipative phase transition marked by the green dashed line in Fig.~\ref{fig:stability}(c), along with this enhanced coupling at large detuning, maximizes the entanglement.}% to $\mathcal{E}^{\rm{max}}_{1|2}=1/3$.
%%
Incidentally, two-mode squeezing can be estimated from the reduced covariance matrix $\tilde{V}^{(k)}_{i|j}$ of the system, in terms of its lowest eigenvalue $\nu^{\rm{min}}$.
The condition $\nu^{\rm{min}}<1/2$ implies that the system is squeezed below the shot noise limit~\cite{Simon1994,Bibak2023}, and is thus squeezed. The variation of the lowest $\nu^{\rm{min}}$ among all stable roots with the effective detuning $\tilde{\Delta}$ of the system is shown in Figs.~\ref{fig:ent_phases}(c)-(d), where it can be directly compared to entanglement. 

\begin{figure}[t]
\includegraphics[width=\columnwidth]{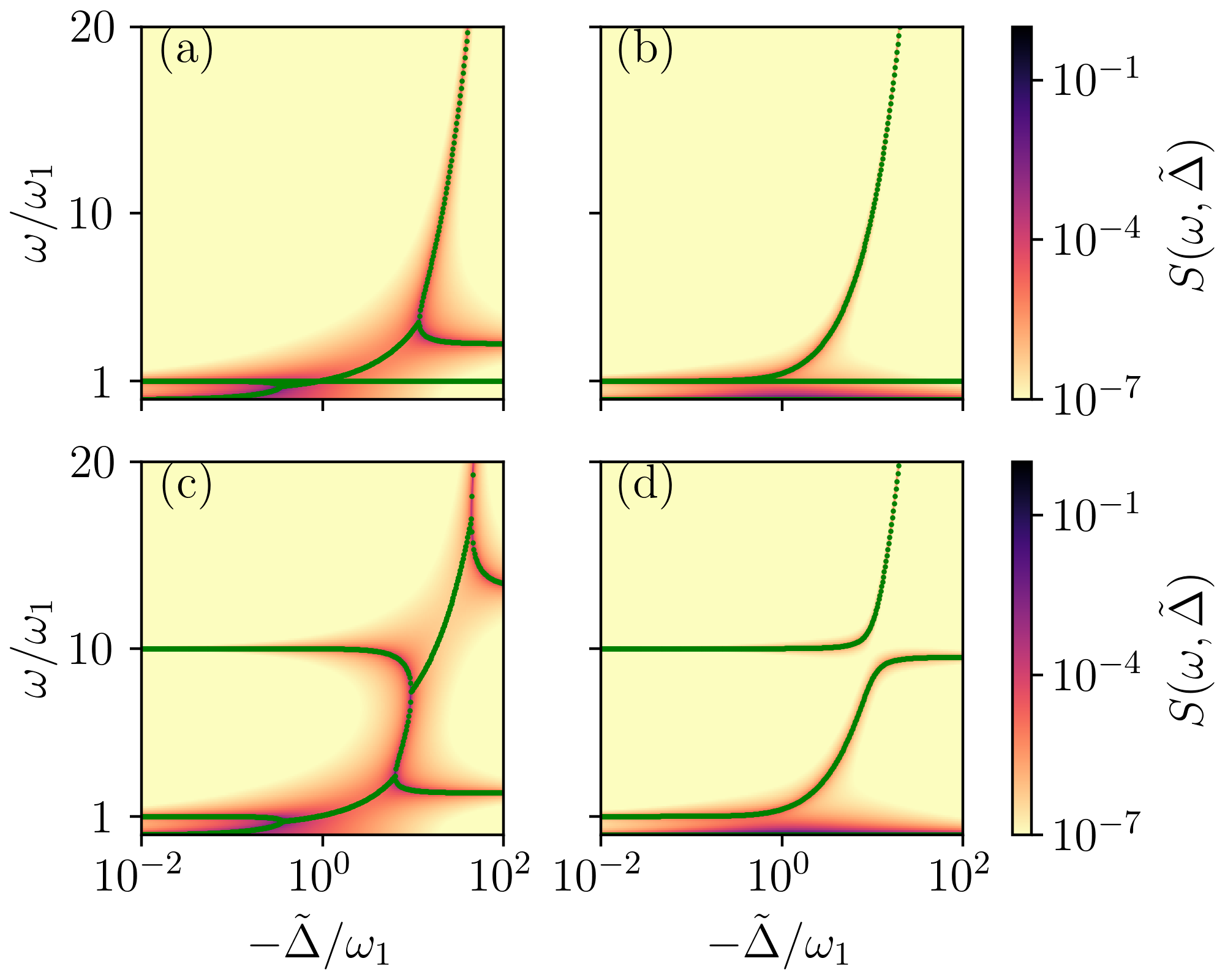}
\caption{Output optical spectrum. The plots (a,b) show the spectrum at $\omega_2=\omega_1$ for the roots $k=1$ and $k=2$, which exhibit level attraction and repulsion, respectively. The bottom (c,d) shows the spectrum at $\omega_2=10~\omega_1$ for $k=1$ and $k=2$, respectively. The green dotted line shows the associated eigenvalues.}
\label{fig:spectral}
\end{figure}

Further qualitative behavior of entanglement and squeezing in the collective modes can be ascertained by the noise spectrum $S(\omega)$ of the optical mode as shown in Fig.~\ref{fig:spectral}. This can be experimentally detected~\cite{Li2022} from the light reflected from the cavity as shown in Fig.~\ref{fig:setup_model}. The calculation of the noise spectrum is shown in Appendix~\ref{app:spectra}. The spectrum exhibits level attraction with eigenvalues merging to form exceptional points (EPs)~\cite{Dembowski2001,Heiss2004} in one of the roots, while level repulsion in another, leading to the formation of entangled hybrid modes of photons and DWs. 
% In other regimes, strong level repulsion at larger detuning leads to decoupling from the cavity mode and stronger DW-DW entanglement. Moreover, the sharp discontinuity in squeezing in Figs.~\ref{fig:ent_phases}(a)-(b), correspond to exceptional points in the {dissipative} phase diagram.
%
% This tells us that under the large detuning regime, both the energy-conserving interaction and the squeezing interaction (second and third terms of Eq.~\eqref{eq:H_ad} respectively) contribute equally with interaction strength $G^{\rm eff}_{j,k}$ between the DW modes $b_j$ and $b_k$. This also shifts the DW oscillator frequency by a self-interaction term $G^{\rm eff}_{j,j}$. %Detailed calculation for deriving steady state covariance matrix analytically for this case is presented in App.~\ref{app:covariance}
% We find that the stable entanglement in both photon-DW modes and DW-DW modes is upper bounded in power by the upper limit of the bistability regime. Thus, this bound below which one gets stable entanglement is given by $\xi^{+}$ from Eq.~\eqref{eq:limits}. The DW-DW entanglement is shown along with this upper bound in Fig.~\ref{fig:ent_phases}. 
%
In Fig.~\ref{fig:spectral}(a) for root $k=1$ we find level attraction in the polaritonic modes formed with a cavity and a collective bright DW mode as they merge just before the resonance $\tilde{\Delta}=-\omega_1$ and separate in the large detuning regime.
%
%
% while, a dark mode remains invariant with the detuning and {has no peak in the spectrum (horizontal green line)}. 
% An exceptional point (EP) is a point in parameter space where two eigenvectors and their eigenvalues coalesce, 
% In this case, there are 
This leads to two second-order exceptional points. First, at the point where the two modes merge {with square root dependency} just before resonance $\tilde{\Delta}=-\omega_1$, and the second, where two modes separate again in the large detuning regime. 
The onset of these two EPs is also associated with the sudden discontinuity in the smallest eigenvalue $\nu^{\rm{min}}_{a|1(2)}$ and the associated two-mode squeezing, as shown in Fig.~\ref{fig:ent_phases}(c). {There also exists a dark mode at $\omega = \omega_{1(2)}$, seen as a horizontal line at $\omega=\omega_1$ in Figs.~\ref{fig:spectral}(a)-(b), which is decoupled from the cavity \cite{Kuzyk2017}, shows no spectral peaks, and remains invariant with the detuning.}  

{A method to understand the origin of EPs is to consider an effective non-hermitian Hamiltonian. This is obtained by rewriting the linearized QLEs in the form of time-dependent Schr\"{o}dinger-like equation for the linearized modes $i\partial_t u^{(k)}=\mathcal{H}u^{(k)}$, neglecting the input noise terms. The eigenvalues of $\mathcal{H}$ are directly related to the drift matrix eigenvalue set as $\{\lambda_\mathcal{H}\}=i\{\lambda_A\}$. In particular, for the case of $\kappa_a=\kappa_{1(2)}$, we obtain exact analytical solutions for six eigenvalues of $\mathcal{H}$. Four of these represent collective bright modes
% $\lambda^{(k)}_{\pm,j}$,
\begin{align}
    \lambda^{(k)}_{\pm,j}=-i\kappa_{a}\pm\sqrt{\tilde{\Delta}^{(k)~2}_{{+}} + (-1)^j \Theta^2}, \label{eq:eigvals}
\end{align}
where $\Theta^2=\sqrt{ \tilde{\Delta}^{(k)~4}_-  - 8 G^2 \bar{n}_a^{(k)} \tilde{\Delta}^{(k)}\omega_1}$ and {$\tilde{\Delta}^{(k)~2}_{\pm}=(\tilde{\Delta}^{(k)~2} \pm {\omega^2_1})/2$}. 
The other two eigenvalues {$\lambda^D_{\pm}=-i\kappa_1\pm\omega_{1}$} are {for} non-interacting dark modes. The formation of second-order EPs occur when the term $\Theta$ inside the square root in Eq.~(\ref{eq:eigvals}) vanishes, leading to coalescence of eigenvalues and eigenvectors, at $\tilde{\Delta}^{(k)}=-\omega_1(2G g_1\bar{x}_a^{(k)}/\kappa_a)^2$ for $j=1$ branch. When $\kappa_a\neq\kappa_1$, the eigenvalues need to be numerically obtained from the drift matrix in Eq.~(\ref{eq:drift_mat}), however, the dark mode eigenvalues retain the same value $\lambda^D_{\pm}$. While crossing these EPs, there are transitions from a region where the real part of two eigenvalues merge while the imaginary part splits as square root and vice versa. 
% Because of this square root nature of splitting, the associated EP is classified as second order. 
Such sudden transition of eigenvalues leads the squeezed state to jump from one stable root to another, leading to a jump in optimal squeezing we see in Fig.~\ref{fig:ent_phases}(c,d)}.

{Similarly, in Fig.~\ref{fig:spectral}(b) for $k=2$, just close to the resonance $\tilde{\Delta}=-\omega_1$, the bright DW mode gets strong level repulsion and the cavity mode remains close to $\omega=0$ while the dark DW mode again remains invariant for all detunings. Thus, this resonance is associated with the maximum DW-photon entanglement $\mathcal{E}_{a|1(2)}$ as observed in Fig.~\ref{fig:ent_phases}(a)}.
{For $\omega_1\neq\omega_2$, we have two resonances at $\tilde{\Delta}=-\omega_1$ and $-\omega_2=-10\omega_1$. For root $k=1$ as shown in Fig.~\ref{fig:spectral}(c), these two resonances thus result in two regimes of level attraction close to each of these resonances and are thus associated with four second-order EPs, each of which leads to discontinuity in the squeezing $\nu^{\rm{min}}_{a|1}$ and $\nu^{\rm{min}}_{a|2}$ as shown in Fig.~\ref{fig:ent_phases}(d). Similarly, for root $k=2$, we see the level repulsion at each of these resonances leading to maximum photon-DW entanglement $\mathcal{E}_{a|1(2)}$ at the respective resonances $\tilde{\Delta}=-\omega_{1(2)}$}.
{The analysis of the eigenvalues of the drift matrix $\{\lambda_A\}$ and EPs can also be extended to study correlations and entanglement at dissipative phases transitions.
% and to investigate the change from photon-DW to DW-DW entanglement along the bifurcation line. 
This is shown in Appendix~\ref{app:ent}.} 

\begin{figure}[t]
	\includegraphics[width=0.85\columnwidth]{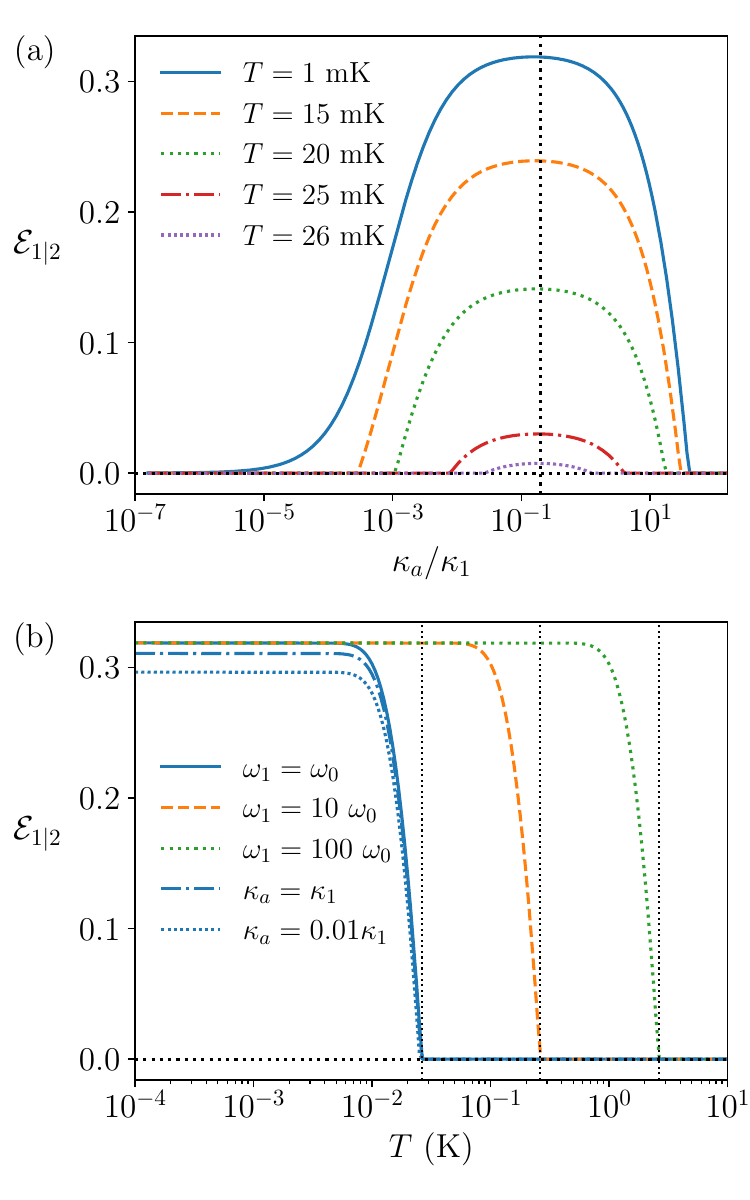}
	\caption{Thermal stability of macroscopic entanglement between the two domain walls. (a) Plot shows entanglement as a function of the ratio $\kappa_a/\kappa_1$ for different temperatures $T$. Here $\omega_{1(2)}=\omega_0$ = 1~GHz, $\tilde{\Delta}=-40~\omega_{1(2)}$ and $\kappa_1=1$~MHz. The vertical black dotted line highlights the maximum entanglement.
    %
    % (a) shows the ratio  $\kappa_a/\kappa_1=0.2$ (black vertical dotted line) can be chosen to get maximum entanglement at any temperature $T<T_c=26.2$mK, here $\omega_{1(2)}$=$\omega_0=1$~GHz, $\tilde{\Delta}=-40~\omega_{1(2)}$ and $\kappa_1=1$~MHz. 
    (b) The plot shows the temperature dependence of $\mathcal{E}_{1|2}$ 
    % at $\tilde{\Delta}=-40\omega_{1(2)}, 
    at fixed $\kappa_a/\kappa_1=0.2$ and varying $\omega_{1(2)}/\omega_0=1,10,100$ with solid blue, orange dashed, and green dotted curves. Now, for fixed $\omega_{1(2)}=\omega_0$, the dissipation is varied, such that $\kappa_a/\kappa_1$ = 1 and 0.01 with blue dot-dashed and dotted curves, respectively. The vertical black dotted lines mark the cutoff temperature $T_c \approx 26$~mK, $0.26$~K, and $2.6$~K for $\omega_{1(2)}$ = 1~GHz, 10~GHz, and 100~GHz, respectively.}
\label{fig:entangle}
\end{figure}

An important aspect related to the robustness of the macroscopic entanglement between the two domain walls is thermal stability. In general, thermal fluctuations at finite temperatures are detrimental to entanglement in any dissipative system operating in the GHz regime. In quantum magnonics, these temperatures are typically in the milli-Kelvin regime~\cite{Tabuchi2015,Bayati2024}, which requires strong cooling efforts, making it experimentally challenging. Figure~\ref{fig:entangle} shows the variation of entanglement between the two DWs with dissipation and temperature. 
Within a range of dissipation values, there exists a cutoff temperature $T_c$ 
% and a range of dissipation ratio $\kappa_a/\kappa_1$, 
beyond which thermal fluctuations prohibit any entanglement between the DWs. 
% However, $T_c$ can be manipulated by changing the key parameters in the system. 
First, we note that altering the loss rates of either the photonic ($\kappa_a$) or %magnonic
{DW} modes and ($\kappa_1$) changes the amount of macroscopic entanglement but not the $T_c$. Remarkably, reducing $\kappa_a$ by a factor of 10 has the same positive effect on entanglement as that achieved by increasing $\kappa_1$ by a factor of 10. As such the entanglement between the DWs is dependent on the ratio $\kappa_a/\kappa_1$, as shown in Fig.~\ref{fig:entangle}(a), with the maximum achieved at 0.2.
However, an increase of the physically relevant cutoff temperature $T_c$ can be achieved by controlling the pinning potentials. This is due to the fact that number of thermal excitations in DW modes that can negate entanglement, is directly dependent on $\hbar\omega/k_BT$. 
In fact, an increase of the DW frequency $\omega_{1(2)}$ by an order of magnitude leads directly to an order of magnitude increase in the cut-off temperature $T_c$.  
As such, as observed in Fig.~\ref{fig:entangle}(b), within realizable experimental parameters, stable, robust entanglement can be sustained between the two domain walls up to {2~K}, using DW frequencies of 100~GHz~\cite{Voto2017-uz,Touko2017}.
% , and as high as {10-Kelvins} for engineered DWs operating in the THz regime~\cite{Hlinka2017}.

% \section{Results and Discussion}\label{sec:res}

% \section{Conclusion and Outlook}\label{sec:con}
\section{Conclusion\label{sec:con}}
% {Need to write this at the end.} 
The generation of stable and robust entanglement in spin textures and magnons is a challenging task and often requires strong nonlinearity and cryogenic temperatures. In this work, we study the dissipative phases of localized domain walls in a chiral cavity and by controlling the pinning potential and the optical driving, steady-state entanglement is generated between the domain walls. Using states close to the transcritical bifurcation line, 
entanglement can be switched between photons and DWs to stable entanglement between the two DWs. For suitable, controlled parameters the entanglement can be sustained up to temperatures as high as a few Kelvins. Given recent experimental developments in the study and control of uniaxial magnetic materials, our study provides technologically relevant insight towards entangling stationary spin textures.

Another significant outcome of our study is the optical spectrum for different values of detuning. In the dispersive regime, for the stable roots, there exist both level attraction and repulsion that highlight different regions where strong photon-DW and DW-DW entanglement are generated. The spectrum also shows the emergence of second-order exceptional points in these systems associated with square root spectral splitting, which can be useful for high-precision, magnetic field sensing~\cite{Wang2021}. 

Finally, we believe the proposed theoretical scheme can be extended to other chiral spin textures, optomechanical and optomagnonic hybrid systems and prove useful in  various quantum information and computation, quantum communication and quantum sensing applications. An interesting platform is domain walls in ferroelectric systems, where photostrictive effects can give rise to strong interactions~\cite{Rubio-Marcos2015}, and pinned domain walls at THz frequencies can be engineered~\cite{Hlinka2017}. This would allow entanglement at temperatures as high as 10~K, which can be relatively easier to setup compared to contemporary cryogenic setups.

% In summary, we have shown that macroscopic entanglement between chiral domain walls can be achieved by coupling them with a single optical cavity mode. We show that a variety of stable dissipative phases exist, some of which support stable entanglement for a large range of parameters. We combine the large detuned dispersive regime with a continuous dissipative phase transition to maximally entangle these spin textures. {We also show that one can identify the exact value of DW mode resonance frequencies by observing the spectrum for different values of optical detuning. An interesting aspect of our system is that one can selectively entangle and disentangle targeted domain walls and photon mode by controlling pinning strengths and optical detunings. We noted that the generated entanglement can be enhanced to an optimal value by increasing detuning around the dissipative phase transition. For optimizing the temperature cutoff for entanglement, we need to increase the pinning strength. Within an experimentally feasible regime, we show that stable DW-DW mode entanglement can survive up to around 2 K theoretically. We also showed the emergence of multiple second-order exceptional points in these systems associated with square root spectral splitting, which can be useful for various precision magnetic field sensing applications.}

\begin{acknowledgments}
We thank K.G. Suresh for valuable discussions. R.G. acknowledges financial support from the Human Resource Development Group of CSIR, India, in the form of a Senior Research Fellowship. H.Y.Y. is supported by the National Key R\& D Program of China (2022YFA1402700) and the National Natural Science Foundation of China (NSFC) (Grant No. 12574132). H.S.D. acknowledges funding from the Science and Engineering Research Board, Department of Science and Technology, India, under the Core Research Grant program (Grant No. CRG/2021/008918) and from the Industrial Research and Consultancy Centre, IIT Bombay (Grant No. RD/0521-IRCCSH0-001). 
\end{acknowledgments}

{\section*{Data Availability}
The codes and data that support the findings of this research article are openly available}~\footnote{The source codes that support the findings of this article are openly available at: \url{https://github.com/qid-iitb/Entanglement_DW}, and include routines, and parameters required to reproduce the key results and figures.}.
% \cite{Gupta2025}, 
% {and include codes, routines, and parameters required to reproduce the key results and figures.}

\appendix

\section{{Quantizing the domain wall and capturing the magneto-optic coupling}\label{app:quantization}}
{As discussed in Sec.~\ref{sec:setup}, under strong pinning potential the domain wall (DW) can collectively be treated as a macroscopic particle.
% Uniaxial ferromagnets support domain walls formation in them, which are easier to manipulate by pinning them with some magnetic nanostructure or deformity \cite{Mironov2016,Jin2017,Lee2023-pin}.
In this section, we consider the canonical quantization of a such a DW inside a cavity, interacting through the inverse Faraday effect~\cite{Proskurin2018}. In the absence of a cavity field, assuming the presence of a pinning potential $U_{\rm pin}(X)$, we write down the effective Hamiltonian describing the collective dynamics of a Bloch DW with its center located at position $X$.
\begin{align}
    \mathcal{H}_{\rm DW}&=\frac{P_X^2}{2m_{\rm eff}} + U_{\rm pin}(X)~~,\nonumber\\
    U_{\rm pin}(X)&=-K_{\rm pin}\rm{sech}^2 \left(\frac{X}{\lambda_{\rm DW}}\right)
\end{align}
where $P_X$ is a canonical momentum obtained from the spin Lagrangian density by shifting to the collective coordinate frame \cite{Proskurin2018}, whose details we have omitted here for simplicity. The domain wall width is taken as $\lambda_{\rm DW}=\sqrt{J/K_{||}}$ where $J$ is the ferromagnetic exchange constant and $K_{||}$ is the in-plane anisotropy field of the material. The pinning potential is typical of the type of a P$\ddot{o}$schl-Teller potential, which under a sufficiently strong pinning strength can be approximated under strong pinning strength as a parabolic potential:
\begin{align}
    U_{\rm pin}(X)&=-K_{\rm pin}\rm{sech}^2(X/\lambda_{\rm DW})= \frac{-K_{\rm pin}}{\left(e^{X/\lambda_{\rm DW}}+e^{-X/\lambda_{\rm DW}}\right)^2}\nonumber\\
    &\simeq \frac{-K_{\rm pin}}{2 + (X/\lambda_{\rm DW})^2}\simeq K_{\rm pin}\frac{X^2}{4\lambda_{\rm DW}^2}
\end{align}
Thus, under this strong pinning regime, we can quantize the oscillator by defining annihilation operators as
$x=x_{\rm zpf}(b + b^\dag)$ and
$p=-im_{\rm eff}\omega x_{\rm zpf} (b -b^\dag)$,
where $x_{\rm zpf}=\sqrt{{\hbar}/{(2m_{\rm eff}\omega)}}$.

It was realized that when a DW oscillator is placed inside a cavity resonator, it can feel the radiation pressure forces from the electromagnetic field analogous to the optomechanical radiation pressure experienced by a suspended mirror in cavity optomechanics. {The origin of such a force here is because of the magneto-optic coupling~\cite{Cottam1986} between the electromagnetic field and the DW, arising from the inverse Faraday effect~\cite{Pitaevskii1961,Pershan1966}. The electric} permittivity tensor gets modulated by the local spin oscillations resulting in the interaction energy
\begin{align}
    \mathcal{H}_{\rm int}&=-\frac{\varepsilon_0}{4}\int\delta\varepsilon_{ij}(\bm{S})\bm{\mathcal{E}}_i(\bm{r},t)\bm{\mathcal{E}}^*_j(\bm{r},t)d^3 r.
\end{align}
Here, the spin density is given as $\bm{S}(\bm{r})$, the complex amplitude of the electric field $\bm{\mathcal{E}}(\bm{r},t)$ is related with original field as $\bm{E}(\bm{r},t)=\rm{Re}(\bm{\mathcal{E}}(\bm{r},t)e^{-i\omega_{a} t})$.
The electric field is quantized as $\bm{\mathcal{E}}(x,t)=-i\sum_{n\lambda}\sqrt{\hbar\omega_n/\epsilon_0 V}\sin(n\pi x/L_{\rm cav})\bm{e}_\lambda a_{n\lambda}$, here $\omega_n$ are the cavity eigenmode frequencies and polarization states in helicity basis are given as $\bm{e}_\lambda=(0,\lambda,-i)/\sqrt{2}$, $\lambda\in\{\pm1\}$. Considering the inverse-Faraday effect induced interaction, the permittivity tensor is given as $\delta\varepsilon_{ij}=if\epsilon_{ijk}S_k$, $\epsilon_{ijk}$ is the Levi-Civita symbol, $f$ depends on the Faraday rotation of the sample $\phi_F$ as $f=2c\phi_F\sqrt{\varepsilon}/\omega_n$ {and $S_k$ is the $k^{\rm{th}}$ component of spin density $\bm{S}(\bm{r}
)$}. {When $N$ DWs are placed inside the cavity, the cavity frequency depends on the electric permittivity, thus modulating it depending on the position of the DWs, $\bm{X}(t)=[X_1(t),X_2(t),...,X_N(t)]^T$. Considering the cavity to be centered at eigen frequency $\omega_n=\omega_a(\bm{X})$, we obtain the coupling as
\begin{align}
    \omega_a(\bm{X})a^\dag a&\simeq \omega_a + \sum_{j=1}^N\frac{\partial \omega_a}{\partial X_j}X_j a^\dag a,~g_j=-\frac{\partial\omega_a(\bm{X})}{\partial X_j}x^j_{\rm{zpf}}\nonumber\\
    \implies\mathcal{H}_{\rm int}&\simeq -\hbar\sum_{j=1}^N g_j (b_j + b_j^\dag) (a^\dag_R a_R - a^\dag_L a_L)
\end{align}
where} $a_R,~a_L$ are the cavity annihilation operators for the right- and left-circularly polarized ($\lambda=\pm1$) states of light. {Both polarized modes exert a perpendicular force on the DW, but in opposite directions. However, the magnitude of
the interaction remains the same and does not affect the system dynamics. 
Experimentally, it is possible to use a chiral cavity resonator, where one of these polarizations can be fully suppressed \cite{Voronin2022}, allowing us to work with single-mode interaction.}
The coupling strength is obtained as $g_j=fS^j_{\rm eff}\omega_a/4$. The dimensionless geometrical factor $S^j_{\rm eff}=x^j_{\rm zpf}A^j_{\perp}/V_c$ depends
% that takes into account of the available 
on the scattering cross-section $A^j_{\perp}$, 
% where the cavity field penetrates it, 
the cavity volume $V_c$ and the zero point fluctuation $x^j_{\rm zpf}=\sqrt{\hbar/2m_j\omega_j}$ of the macroscopic DW, with mass {$m_j=\hbar^2/(K_{\perp}l\lambda_{\rm{DW}})$, where $l$ is the size of unit lattice and $K_{\perp}$ is the anisotropy constant of the material. We note that higher-order terms and the non-linear Kerr effect have been neglected.} 

{We note that there are qualitative similarities and differences between N{\'e}el and Bloch DWs. The interaction between the DWs and the chiral cavity via the magneto-optic coupling is quite similar for both and can be modeled as in Sec.~\ref{sec:setup}. However, the direction of the radiation pressure is different for N{\'e}el walls, as the axis of rotation of spins is perpendicular to the plane of the DW. 
% \textcolor{blue}{both and can be modelled with optomechanical radiation pressure as we have taken, but in a different direction because in N{\'e}el walls,  the axis of rotation of spins across the walls is perpendicular to the plane of DWs.} 
Importantly, higher DW oscillation frequencies can be obtained in N{\'e}el DWs
% , particularly in antiferromagnetic systems
\cite{Iyaro2021}. However, the coupling is significantly weaker, and requires specialized geometry that allows for Dzyaloshinskii-Moriya interactions.}

% {We note that there are qualitative similarities and differences between N{\'e}el and Bloch DWs. The interaction between the DWs and the chiral cavity via the inverse Faraday effect is similar for both. In fact, higher oscillation frequencies can be obtained in N{\'e}el DWs, particularly in antiferromagnetic systems~\cite{Iyaro2021}. However, the magneto-optic coupling is significantly weaker, and requires specialized geometry that allows for the presence of Dzyaloshinskii-Moriya interactions.}

\section{{System parameters}\label{app:parameters}}

% {For a typical THz cavity with $\omega_a\approx1$THz, one can achieve $S^j_{\rm{eff}}\approx10^{-6}$, which for a typical YIG material with $K_{\perp}\approx0.008K$, pinning strength $K_{\rm{pin}}\approx0.36 K$ for $m_j\approx10^{-27}$kg, $\omega_{j}=1$GHz, $x_{\rm{zpf}}\approx10$nm and $\lambda_{DW}\approx100$nm, $\alpha_G=3\times10^{-5}$ whcih gives $\kappa_j\approx1$MHz. The iron garnet materials with ytterbium and bismuth would be the most suitable for entangling domain walls because of their higher Faraday rotation and lower Gilbert damping rates. The typical cross-sectional dimension of the ferromagnetic strips would be a width $w\approx10\mu$m and height $h=50\mu$m, placed inside a 100$\mu$m cavity size.}

{The physical parameters used to obtained the desired oscillation frequency of the Bloch domain walls and the magneto-optic coupling strength can be achieved in a physical implementation using iron garnets~\cite{Nakamura2024} with ytterbium (YIG) or bismuth and cerium substituted YIG variants~\cite{Fakhrul2019,Hayashi2024}), which are attractive for magneto-optic based interactions and entanglement, because they exhibit large Faraday rotation with very low magnetic damping. In fact materials such as Bi:YIG and Ce:YIG films show Faraday rotation that can be an order(s) of magnitude larger than plain YIG~\cite{Sekhar2012}.}

{Let us consider a chiral cavity with frequency $\omega_a\approx1$~THz and a typical YIG material with anisotropy constant $K_{\perp}\approx0.008$~K, domain wall $\lambda_{DW}\approx100$~nm and unit lattice cell $l\approx1$~nm. The effective mass of the DW oscillator is $m_j=\hbar^2/(K_{\perp}l\lambda_{\rm{DW}})\approx10^{-27}$~kg. For a pinning strength, $K_{\rm{pin}}\approx0.36$~K the DW oscillator frequency is $\omega_{j}\approx1$~GHz, which leads to the zero point fluctuation $x^j_{\rm zpf}=\sqrt{\hbar/2m_j\omega_j}\approx 10~$nm. 
% For the above $m_j$ and $x^j_{\rm zpf}$, and at low temperatures, the DW is operating in the quantum regime, 
The above $m_j$ and $x^j_{\rm zpf}$, represent quantized excitations above classical spin texture, which can then be modeled as quantum harmonic oscillator as shown in Sec.~\ref{sec:setup}. 
For geometrical factor $S^j_{\rm{eff}}\approx10^{-6}$, the single photon coupling is of the order of $1~$MHz, which is the interaction strength necessary to observe the entanglement properties. 
% $S^j_{\rm eff}$ is dependent on the scattering cross-section.
%%
Other key properties associated with YIG materials such as Faraday rotation, $\phi_F = 240^\circ cm^{-1}$, and low damping rate of the order of $10^{-5}$~\cite{Proskurin2018}. The typical cross-sectional dimension of the ferromagnetic strips would be a width $w\approx10\mu$m and height $h=50\mu$m, which is then placed inside a 100$\mu$m cavity size.
An potential extension of the design is the possibility to arrange a series of these cavities and strips to form an magneto-optical array~\cite{Shang2024}.}

\section{Linearizing around steady states \label{sec:appA}}

We discuss the linearization of the system around its steady state, as mentioned in Sec.~\ref{sec:snp}}. {We take the cavity loss rate $\kappa_a$ and dissipation rate in the domain wall modes as $\kappa_j=\alpha_G \omega_j/\sqrt{K^j_{\rm{pin}}l/2K_{\perp}\lambda_{\rm{DW}}}$, where $\alpha_G$ is the Gilbert damping rate~\cite{Tatara2008,Proskurin2018}. These we use for writing quantum Langevin equations (QLEs)}
% Here we will begin with Eq.~\ref{eq:H'} and discuss its linearization around the steady state as mentioned in Sec.~\ref{sec:snp}}. 
%The quantum Langevin equations (QLEs)
for the Hamiltonian in Eq.~(\ref{eq:H'}) are
\begin{align}
    \dot{a}&=(i\Delta_a -\kappa_a)a - i\sum_j g_j (b_j + b^\dag_j)a + \xi + \sqrt{\kappa_a}a^{\rm{in}},\nonumber\\
    \dot{b}_j&=-(i\omega_j+\kappa_j)-ig_j a^\dag a + \sqrt{2\kappa_j}b^{\rm{in}}_j.
    \label{eq:Full_QLEs}
\end{align}
These are obtained by writing the Heisenberg equations and adding the input noise operators $a^{\rm{in}},b^{\rm{in}}_j$ for the photon and DW modes, respectively. These carry the thermal and quantum noise from the input port and follow the following correlations~\cite{Aspelmeyer2014}:
\begin{align}
    &\langle a^{\rm{in}}(t) a^{\rm{in}\dag}(t')\rangle=\left(1+n_a^{\rm{th}}\right)\delta(t-t'),\\
    &\langle a^{\rm{in}\dag}(t) a^{\rm{in}}(t')\rangle=n_a^{\rm{th}}\delta(t-t'),\\
    &\langle b^{\rm{in}}_j(t) b^{\dag \rm{in}}_j(t')\rangle=\left(1+n_j^{\rm{th}}\right)\delta(t-t'),\\
    &\langle b^{\dag \rm{in}}_j(t) b^{\rm{in}}_j(t')\rangle=n_j^{\rm{th}}\delta(t-t'),\\
    &n^{\rm{th}}_a=(e^{\hbar\omega_a/k_B T}-1)^{-1},~n^{\rm{th}}_j=(e^{\hbar\omega_j/k_B T}-1)^{-1}.
\end{align}
These noise fluctuations drive the system along with the coherent photon pumping rate $\xi$. These noise operators have zero mean, $\langle a^{\rm{in}}\rangle=\langle b^{\rm{in}}_j\rangle=0$, similarly by defining $\langle a\rangle=\alpha,~\langle b_j\rangle=\beta_j$, we can write the meanfield representation of Eq.~\eqref{eq:Full_QLEs} as
\begin{align}
    \dot{\alpha} &=\left[i\left(\Delta_{a} - \sum_{j=1}^{N} g_{j}(\beta^{*}_{j}+\beta_{j})\right) - \kappa_{a}\right]\alpha+\xi,\nonumber\\
    \dot{\beta_{j}}&=-(\kappa_{j}+i\omega_{j})\beta_{j} - i g_{j}|\alpha|^2.
\label{eq:MF_EQNs}
\end{align}
Now, as the system drifts towards the steady state as $t\rightarrow\infty$, we expect these time derivatives to vanish which gives the steady state values as a set of coupled equations. Solving the for steady state gives us the following solution:
\begin{align}
    \bar{\alpha} &=\frac{i\xi}{\left(\Delta_{a} - \sum_{j=1}^{N} g_{j}(\bar{\beta}^{*}_{j}+\bar{\beta}_{j})\right) + i\kappa_a},\\
    \bar{\beta}_{j} &=\frac{-ig_{j}|\bar{\alpha}|^2}{i\omega_j+\kappa_j}.
\label{eq:SS_soln}
\end{align}
These can be further simplified if we substitute $\bar{\alpha}=\sqrt{\bar{n}_\alpha}e^{i\bar{\phi}_\alpha},\bar{\beta}_j=\sqrt{\bar{n}_j}e^{i\bar{\phi}_j}$ and demanding $\bar{n}_\alpha,\bar{n}_j\geq0$:
\begin{align}
    &\bar{n}_\alpha\left[\left(\Delta_a +\Omega\bar{n}_\alpha\right)^2 + \kappa_a^2\right] -\xi^2=0,~\textrm{where}\\
    &\bar{\phi}_\alpha=\arctan{\left(\frac{\kappa_a}{\Delta_a+\Omega\bar{n}_\alpha}\right)},~\Omega=\sum_{j=1}^{N}\frac{2g^2_j\omega_j}{\omega^2_j+\kappa^2_j},\\
    &\bar{n}_{j} =\frac{g^2_{j}\bar{n}^2_\alpha}{\omega^2_j+\kappa^2_j},~\textrm{and}~\bar{\phi}_j=\arctan{\left(\frac{\kappa_j}{\omega_j}\right)}.
\label{eq:SS_soln_cubics}
\end{align}
This will have either one or three solutions for mean photon number, we label them by $\bar{n}^{(k)}_\alpha,~k\in\{1,2,3\}$. Now, we shift the Hamiltonian around these steady state values for each root~\cite{Bibak2023}, i.e., transformation $a\rightarrow\alpha+a,b_j\rightarrow\beta_j+b_j$, and
rewrite $H^{(k)}$ (up to two operator terms)  
% in terms of these solutions 
% similar to \cite{Bibak2023} and 
to obtain:
\begin{align}
    H^{(k)}&=-\Tilde{\Delta}^{(k)}a^{\dag}a + \sum_{j=1}^{N}\omega_j b_j^{\dag}b_j,\nonumber\\
    &+\sum_{j=1}^{N} g_j\left(\alpha^{(k)} a^{\dag} + \alpha^{*(k)} a\right)\left(b_{j}+b_j^{\dag}\right).
    % H^{(k)}&=H^{(k)}_2 + H_3.
\end{align}
Now at this stage, a simplification is possible if we make $\vert \alpha\vert\gg 1$, which enables us to drop the cubic term in fluctuations, leading to a simplified quadratic Hamiltonian which leads to Gaussian dynamics. We have ensured that this condition is maintained for all our results.
%In the coming sections we will focus on results for two such DWs but we have verified the results remains qualitatively same for N number of DWs and one can select any of the two such strips by lowering the pinning field at the other DWs thus effectively decoupling them from the dynamics.

\section{Adiabatic Elimination \label{app:adia}}

The linearized Langevin equations can be written as
\begin{align}
    &\delta\dot{a}=(i\tilde{\Delta} - \kappa_a)\delta a - i\alpha\sum_{j=1}^{N} g_{j}(\delta b^\dag_j+\delta b_j) + \sqrt{2\kappa_a}a^{\rm in},\nonumber\\
    &\delta \dot{b}_j=-(\kappa_{j}+i\omega_{j})\delta b_j - i g_{j}(\alpha^*\delta a + \alpha\delta a^\dag) + \sqrt{2\kappa_j}b_j^{\rm in}.
\label{eq:lin_EQNs}
\end{align}
Now, since we are interested in large detuning regime $(\omega_j,\tilde{\Delta}\gg\alpha g_j,~\forall j$), we define slowly evolving variables $\delta\tilde{a}=e^{(-i\tilde{\Delta} + \kappa_a)t}\delta a,~\tilde{b}_j=e^{(i\omega^{\rm eff}_j + \kappa_j)t}b_j$, where we have shifted the mechanical frequency and damping to accommodate for the collective optical spring effect. The cavity part in this frame is obtained as
\begin{align}
    \delta \dot{\tilde{a}}&=-i\alpha\sum_{j=1}^{N} g_{j}\left[\delta \tilde{b}^\dag_j e^{i(\omega^{\rm eff}_j-\tilde{\Delta})t + (\kappa_a-\kappa^{\rm eff}_j)t}\right. \nonumber\\
    &\left. + \delta \tilde{b}_j e^{-i(\omega^{\rm eff}_j+\tilde{\Delta})t + (\kappa_a-\kappa^{\rm eff}_j)t}\right]+\sqrt{2\kappa_a}\tilde{a}^{\rm in}.
\end{align}
We solve the above equation by taking the slowly varying operators outside of the integration over time from $-\infty$ to zero assuming $(\kappa_a\geq\kappa_j)$ to obtain
\begin{align}
    \delta\tilde{a}&=-i\alpha\sum_{j=1}^{N} g_j\left[\frac{\delta\tilde{b}^\dag_j e^{i(\omega^{\rm eff}_j-\tilde{\Delta})t + (\kappa_a-\kappa^{\rm eff}_j)t}}{i(\omega^{\rm eff}_j-\tilde{\Delta}) + \kappa_a-\kappa^{\rm eff}_j}\right.\nonumber\\
    &\left. + \frac{\delta\tilde{b}_j e^{-i(\omega^{\rm eff}_j+\tilde{\Delta})t + (\kappa_a-\kappa^{\rm eff}_j)t}}{-i(\omega^{\rm eff}_j+\tilde{\Delta}) + \kappa_a-\kappa^{\rm eff}_j}\right],~\textrm{where}\\
    \delta a&=-i\alpha\sum_{j=1}^{N} g_j\left[\frac{\delta b^\dag_j}{i(\omega^{\rm eff}_j-\tilde{\Delta}) + \kappa_a-\kappa^{\rm eff}_j}\right.\nonumber\\
    &\left. + \frac{\delta b_j}{-i(\omega^{\rm eff}_j+\tilde{\Delta}) + \kappa_a-\kappa^{\rm eff}_j}\right],\\
    \delta a&=\sum_{j=1}^{N}\left(G^{-}_{j} \delta b^\dag_j+G^{+}_{j} \delta b_j\right).\label{eq:D5}
\end{align}
Substituting Eq.~\eqref{eq:D5} back to the linear equation Eq.~(\ref{eq:lin_EQNs}):
\begin{align}
    \delta \dot{b}_j&=-(\kappa_{j}+i\omega_{j})\delta b_j - \sum_{k=1}^N i\left(G_{j,k}\delta b^\dag_k - G^*_{j,k}\delta b_k\right),\nonumber\\
    G_{jk}&=\frac{2\tilde{\Delta}|\alpha|^2 g_j g_k}{(i\omega^{\rm eff}_k + \Delta \kappa_k)^2 + \tilde{\Delta}^2},~\Delta\kappa_k=\kappa_a-\kappa^{\rm eff}_k. \\
    \delta \dot{b}_j&=-(\kappa^{\rm eff}_{j}+i\omega^{\rm eff}_{j})\delta b_j - i G_{jj}b^\dag_j \nonumber\\
    &-\sum_{k}^{k\neq j} i\left(G_{j,k}\delta b^\dag_k - G^*_{j,k}\delta b_k\right), \nonumber\\
    \omega^{\rm eff}_j&=\omega_j-\text{Re}(G^*_{jj}),~\kappa^{\rm eff}_j=\kappa_j+\text{Im}(G^*_{jj}).\label{eq:eff_shifts}
\end{align}
By solving the Eq.~(\ref{eq:eff_shifts}), one can obtain the shifted damping and frequency induced by the collective optical spring effect. The quadratic interaction Hamiltonian $H_{\rm eff}$ after adiabatic elimination is
\begin{align}
    H^{\rm ad}_{\rm int} &=\sum_{j,k}\left(G_{j,k}\delta b_j\delta b^\dag_k + G^*_{j,k}\delta b^\dag_j\delta b_k\right)\nonumber\\
    &+\sum_{j,k}\left(G_{j,k}\delta b^\dag_j\delta b^\dag_k + G^*_{j,k}\delta b_j\delta b_k\right),
\end{align}
while the non-interacting Hamiltonian is given by
\begin{align}
    H^{\rm ad}_0&=-\tilde{\Delta}\sum_{j,k}\left(G^{+*}_j G^{+}_k\delta b^\dag_j b_k + G^{-*}_jG^{-}_k\delta b_j b^\dag_k\right)\nonumber\\
    &-\tilde{\Delta}\sum_{j,k}\left(G^{+*}_j G^{-}_k\delta b^\dag_j b^\dag_k + G^{-*}_j G^{+}_k\delta b_j b_k\right)\nonumber\\
    &+\sum_j \omega_j \delta b^\dag_j \delta b_j.
\end{align}
Collecting and identifying relevant quadratic terms in the full adiabatic Hamiltonian $H^{\rm ad}=H^{\rm ad}_0+H^{\rm ad}_{\rm int}$, we get:
\begin{align}
    H^{\rm ad}&=\sum_j\left[\omega^{\rm eff}_j\delta b^\dag_j b_j + \frac{1}{2}\left(G_{jj}\delta b^{\dag 2}_j + G^*_{jj}\delta b^2_j\right)\right]\nonumber\\
    &+\sum^{j\neq k}_{j,k} \left(\nu^+_{j,k}\delta b^\dag_j \delta b_k + \nu^-_{j,k}\delta b_j \delta b^\dag_k\right)\nonumber\\
    &+\sum^{j\neq k}_{j,k}\left(\mu_{j,k}\delta b^\dag_j\delta b^\dag_k + \mu^*_{j,k}\delta b_j b_k\right),~\textrm{where}\\
    \omega^{\rm eff}_j&=\omega_j + 2\text{Re}(G_{jj}) - \tilde{\Delta}\left(|G^+_j|^2 + |G^-_j|^2\right),\nonumber\\
    \nu^-_{j,k}&=G_{j,k}-\tilde{\Delta}G^{-*}_jG^-_k,~\nu^+_{j,k}=G^*_{j,k}-\tilde{\Delta}G^{+*}_jG^+_k,~\rm{and}\nonumber\\
    \mu_{j,k}&=G_{j,k}-\tilde{\Delta}G^{+*}_jG^-_k.
\end{align}
Using this total adiabatic Hamiltonian, we obtain the Langevin equations of motion:
\begin{align}
    i\delta \dot{b}_j&=(\omega^{\rm eff}_j - i\kappa_j)\delta b_j +G_{jj}\delta b^\dag_j + \sum_k^{k\neq j} \left(\nu^+_{jk}\delta b_k + \mu_{jk}\delta b^\dag_k\right),\nonumber\\
    i\delta \dot{b}^\dag_j&=(-\omega^{\rm eff}_j - i\kappa_j)\delta b^\dag_j -G^*_{jj}\delta b_j - \sum_k^{k\neq j} \left(\nu^-_{jk}\delta b^\dag_k + \mu^*_{jk}\delta b_k\right).
\label{eq: ad_lineqs}
\end{align}
We note that this gives different strengths for beam splitter interaction and for two mode squeezing interaction, which is different from what one obtains from very large detuning regime $(\tilde{\Delta}\gg \omega_j\geq |\alpha g_j|)$. For such a regime a great simplification is possible using
\begin{align}
    G^{\rm eff}_{j,k}&\simeq\frac{2|\alpha|^2 g_j g_k}{\tilde{\Delta}},~G^{\pm}_j\simeq\frac{\alpha g_j}{\tilde{\Delta}},~\nu_{j,k}=\mu_{j,k}\simeq\frac{|\alpha|^2 g_j g_k}{\tilde{\Delta}}.\nonumber
\end{align}
% Similarly, we can obtain the effective input cavity noise coming with mechanical modes:
% \begin{align}
%     \delta b_j
% \end{align}
\begin{figure*}
    \centering
    \includegraphics[width=\linewidth]{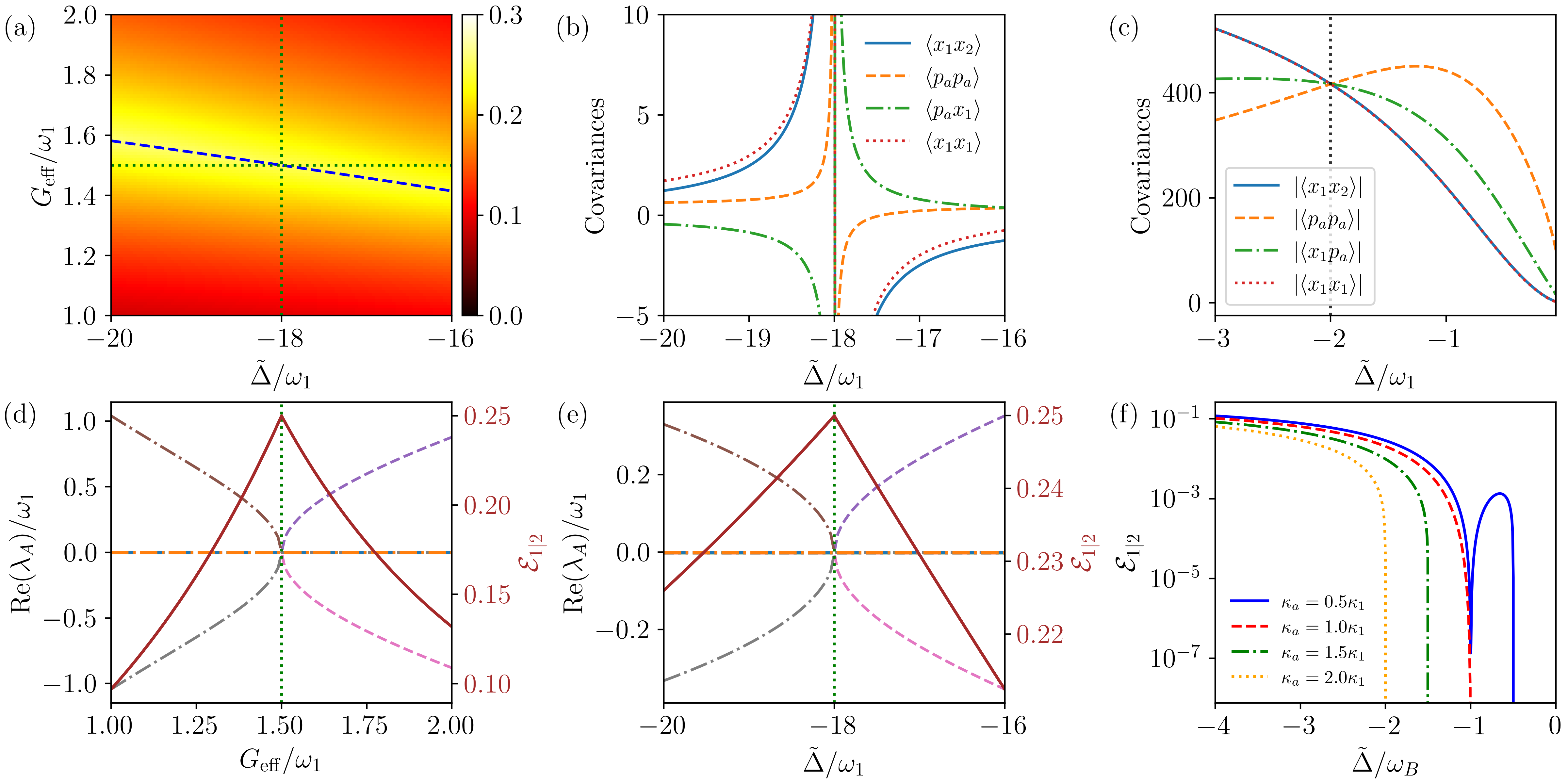}
    \caption{{Entanglement near dissipative phase transition. (a) The plot shows the entanglement variation with effective coupling $G_\text{eff}/\omega_1$ and detuning $-\tilde{\Delta}/\omega_1$, for a small region of Fig.~\ref{fig:stability}(c). The blue dashed line is the bifurcation line. (b) The change in local and intermode covariance of photon and DW modes at $G_\text{eff}/\omega_1 \approx 1.5$. (d) and (e) show the eigenvalues of the drift matrix $\{\lambda^{(k)}_A\}$ as it varies with effective coupling and detuning, respectively. These are changes across the green-dotted line in (a). The plots in (c) and (f) show the change in intermode covariance and entanglement, respectively,  in the vicinity of the bifurcation line (green dashed line in Fig.~\ref{fig:stability}(a)) as the detuning is $-\tilde{\Delta}$ is increased. The plots highlight the critical detuning for onset of DW-DW entanglement, which is equal to the ratio of photon and DW mode loss rate $\kappa_a/\kappa_1$.}}
    \label{fig:Ent_origin}
\end{figure*}

This is what one can also obtain from the second-order perturbation theory assuming that cavity fluctuations remain in a vacuum state in the zeroth order \cite{Schmidt2012}, however, this leads to symmetric coefficients. Our results are valid even in a regime where resonances are dominant and 
where these coefficients are quite different, and they extend to the large detuning regime.

\vspace{-.5cm}
{\section{Correlations near the dissipative phase transition~\label{app:ent}}}

{To qualitatively understand the physical origin of entanglement across the bifurcation line, we investigate the  eigenvalues of the drift matrix $\lambda_A$, similar to the analysis done in Sec.~\ref{sec:entangle}, and also the correlations between the different modes.
While the former tells us about the nature of the phase transition that leads to divergence of mode covariances, the latter highlights the specific correlations that build up between the mode quadratures.
Figure~\ref{fig:Ent_origin}(a), shows the entanglement in a specific region of the phase space, close to the continuous dissipative phase transition line (blue-dashed line).
The dissipative phase transition occurs when the steady state roots $k=0$ and $k=2$ exchange their stability. 
Figures~\ref{fig:Ent_origin}(d,e) shows that the real part of $\lambda_A$ approaches zero as one gets close to the critical line by either changing the effective coupling $G_\text{eff}$ (Fig.~\ref{fig:Ent_origin}(b)) or the dispersion $\tilde{\Delta}$ (in Fig.~\ref{fig:Ent_origin}(c)). The square root shaped bifurcation gives rise to an exceptional point with a significant build up DW-DW entanglement. In fact, the entire  transition line is an exceptional line~\cite{Ding2022}, where a pair of eigenvalues coalesce to zero at each point, which results in large correlations. 
For instance, Fig.~\ref{fig:Ent_origin}(b) shows the divergence of the intermode covariances $\langle x_1 x_2\rangle$ and $\langle p_a x_1\rangle$, and the local mode fluctuations $\langle x_1x_1\rangle$ and $\langle p_a p_a\rangle$. The covariances show that the quadratures $x_1$ and $x_2$ are correlated and comparable to the local fluctuations of $x_1$ and  $x_2$, which entangles the two DW modes. Moreover, the quadratures $p_a$ and $x_1$ are anticorrelated and comparable with $\langle p_a p_a\rangle$, leading to entanglement between photon and DW modes, as evident from calculations of logarithmic negativity. }

{Another important observation is the onset of DW-DW entanglement along the bifurcation line. Analyzing the covariances in Fig.~\ref{fig:Ent_origin}(c), we see that the onset of DW-DW entanglement (Fig.~\ref{fig:Ent_origin}(f)) corresponds to the point $-\tilde{\Delta}/\omega_1 = 2$, where $\langle x_1 x_2\rangle = \langle x_1 x_1\rangle > \langle p_a x_1 \rangle > \langle p_a p_a \rangle$. As such, the DW-DW entanglement increases as photo-DW entanglement decreases. 
 %%
% As the dispersion is increased i.e., $-\tilde{\Delta}>\omega_1$ and the system moves away from resonance ($-\tilde{\Delta}\approx 0$), it transitions from maximal photon-DW to DW-DW entanglement. 
Interestingly, Fig.~\ref{fig:Ent_origin}(f)) shows that DW-DW entanglement along the transition line becomes non-zero at $-\tilde{\Delta}/\omega_1 = \kappa_a/\kappa_1$, where the effective dispersion $|\Delta/\omega_1|$ starts becoming larger than the ratio of photon and DW loss rates.} \\

\section{Output optical spectrum\label{app:spectra}}
{In this section, we 
% will bridge to the experiments by connecting 
connect the dissipative phases associated with DW-DW entanglement to the experimentally observable optical spectrum~\cite{Li2022}. This is measured by collecting the reflected output light and passing it through a spectrum analyzer as shown in Fig.~\ref{fig:setup_model}}.

For this, here we provide a simple compact matrix method to calculate the spectrum for any quadrature set. The linearized quantum Langevin equations can be written as a matrix and expressed in the frequency domain:
\begin{align}
    \dot{u}&=Au + D_0 u^{\rm{in}},~\rm{where}~\eta=D_0u^{\rm{in}},\nonumber\\
    u(\omega)&=-(A+i\omega I)^{-1}D_0u^{\rm{in}}(\omega).
\end{align}
Then we combine it with input-output formalism to write the output operators in terms of the input noise operators
\begin{align}
    u^{\rm{out}}(\omega)&=D_0 u(\omega) - u^{\rm{in}}(\omega),\nonumber\\
    u^{\rm{out}}(\omega)&=-\left[I+D_0(A+i\omega I)^{-1}D_0\right]u^{\rm{in}}(\omega)=\mathcal{T}(\omega)u^{\rm{in}}(\omega).
\end{align}
% \begin{align}
%%
The output spectral function $S_{i,j}(\omega)$ for correlations between $u^{\rm out}_i,u^{\rm out}_j$ quadrature operators can be written as
\begin{align}
    S_{i,j}(\omega)&=\frac{1}{2\pi}\int^{\infty}_{-\infty}\langle u^{\rm{out}}_i(\omega) u^{\rm{out}}_j(\omega')\rangle d\omega',\nonumber\\
    S_{i,j}(\omega)&=\frac{1}{2\pi}\sum_{k,l}\int^{\infty}_{-\infty}\mathcal{T}_{i,k}(\omega)\mathcal{T}_{j,l}(\omega')\langle u^{\rm{in}}_k(\omega) u^{\rm{in}}_l(\omega')\rangle d\omega'.
\end{align}
All the expectation values can be combined in the form of a matrix $\chi$: $\langle u^{\rm{in}}_k(\omega) u^{\rm{in}}_l(\omega')\rangle=2\pi\chi_{k,l}(\omega)\delta(\omega + \omega')$. {The matrix elements $\chi_{k,l}$ can be obtained by using the Fourier transformation of the input noise correlations and converting them into quadrature variables
\begin{align}
    \langle x^{\rm{in}}_k(\omega)x^{\rm{in}}_l(\omega')\rangle&= 2\pi\delta(\omega+\omega')\frac{(1+2n^{\rm{th}}_k)}{2}\delta_{k,l}, \nonumber\\
    \langle p^{\rm{in}}_k(\omega)p^{\rm{in}}_l(\omega')\rangle&= 2\pi\delta(\omega+\omega')\frac{(1+2n^{\rm{th}}_k)}{2}\delta_{k,l}, \nonumber\\
    \langle x^{\rm{in}}_k(\omega)p^{\rm{in}}_l(\omega')\rangle&= 2\pi\delta(\omega+\omega')\frac{i}{2}\delta_{k,l}, \nonumber\\
        \langle p^{\rm{in}}_k(\omega)x^{\rm{in}}_l(\omega')\rangle&= -2\pi\delta(\omega+\omega')\frac{i}{2}\delta_{k,l},
\label{eq:freq_corr}
\end{align}
where $k,l\in\{a,1,2\}$. Using this we get
\begin{align}
    \chi=\begin{pmatrix}
        \frac{1+2n^{\rm{th}}_a}{2} & \frac{i}{2} & 0 & 0 & 0 & 0\\
        \frac{-i}{2} & \frac{1+2n^{\rm{th}}_a}{2} & 0 & 0 & 0 & 0\\
        0 & 0 & \frac{1+2n^{\rm{th}}_1}{2} & \frac{i}{2} & 0 & 0\\
        0 & 0 & \frac{-i}{2} & \frac{1+2n^{\rm{th}}_1}{2} & 0 & 0\\
        0 & 0 & 0 & 0 & \frac{1+2n^{\rm{th}}_2}{2} & \frac{i}{2}\\
        0 & 0 & 0 & 0 & \frac{-i}{2} & \frac{1+2n^{\rm{th}}_2}{2}
    \end{pmatrix},
\end{align}}
and the expression of the spectral function simplifies to
\begin{align}
    S_{i,j}(\omega)&=\sum_{k,l}\int^{\infty}_{-\infty}\mathcal{T}_{i,k}(\omega)\mathcal{T}_{j,l}(\omega')\chi_{k,l}(\omega)\delta(\omega + \omega') d\omega',\nonumber
\end{align}
\begin{align}
    S_{i,j}(\omega)&=\sum_{k,l}\mathcal{T}_{i,k}(\omega)\chi_{k,l}(\omega)\mathcal{T}_{j,l}(-\omega),~\rm{and}\nonumber\\
    \mathcal{S}(\omega)&=\mathcal{T}(\omega)\chi(\omega)\mathcal{T}^T(-\omega).
    \label{eq:simplified_spectra}
\end{align}
The optical output spectrum is then obtained by summing over both the position and momentum cavity quadrature correlations. 
{Since the spectral function consists of noise correlations and expectation of quadrature variables in Eq.~\ref{eq:freq_corr}, it must satisfy the relation 
$S_{x_a,x_a}(\omega)+S_{p_a,p_a}(\omega)\geq 1$, which arises from the uncertainty relation of the noise spectrum $S_{x_a,x_a}(\omega)S_{p_a,p_a}(\omega)\geq1/4$~\cite{Braginsky1992, Clerk2010}.
%
% are given by the integral of relations in Eq.~\ref{eq:freq_corr}, the spectrum satisfies the inequality $S_{x_a,x_a}(\omega)+S_{p_a,p_a}(\omega)\geq 1$, which can be shown directly by applying the inequality of arithmatic mean and geometric mean to $S_{x_a,x_a}(\omega)S_{p_a,p_a}(\omega)\geq1/4$, which is the Robertson-Schrodinger uncertainty relation \cite{Robertson1929}. It is extended to the noise spectrum \cite{Clerk2010} to continous variable systems at each frequency point $\omega$, and can be even more generalized from the Maccone-Pati uncertainity relations\cite{Maccone2014}, if one needs a tighter bound. 
%
However, to achieve better contrast while plotting the spectrum, we define $S(\omega)=S_{x_a,x_a}(\omega)+S_{p_a,p_a}(\omega)-1$, such that $S(\omega)\geq 0$. This is the quantity plotted as the output optical spectrum in Fig.~\ref{fig:spectral}.}

\begin{figure}[t]
\centering
    \includegraphics[width=0.9\columnwidth]{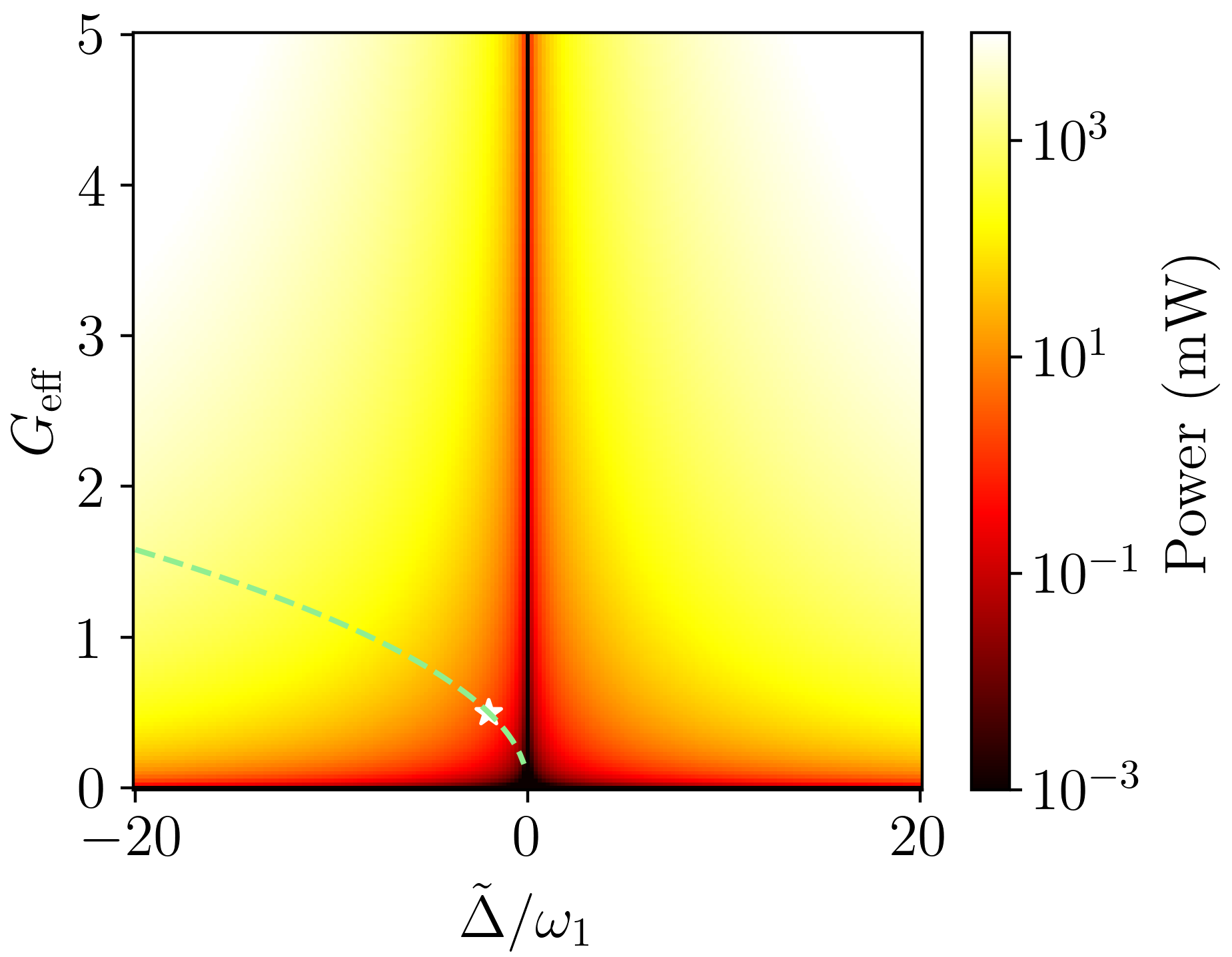}
    \caption{{Pump power as a function of effective coupling $G_{\rm{eff}}$ and detuning $\tilde{\Delta}$. The light green dashed line is the phase transition line where the DW-DW entanglement is maximum, and the star marks the onset of entanglement as $\tilde{\Delta}$ decreases below resonance.}}
    \label{fig:power_var}
\end{figure}

{\section{Optical driving frequency and power}}

{We analyze the optical frequency and power of the driving field necessary
to achieve the desired DW-DW entanglement. 
% {The power in our analysis is given by the relation $\hbar\omega_a|\xi|^2/2\kappa_a$.}
%
Figure~\ref{fig:power_var} shows the pump power, given by the relation  $\hbar\omega_a|\xi|^2/2\kappa_a$, required to reach the different steady state phases in Fig.~\ref{fig:stability}. 
% of the manuscript.
The onset of entanglement between the two domain walls along the transition line (shown by a star in Fig.~\ref{fig:power_var}) is achieved with only a few mWs of power. For maximum attainable values of DW-DW entanglement 
% Thus, one can achieve DW-DW entanglement with even a few mWs of laser power, while maximal values
a pump power of around 100~mW, operating in the THz regime, is needed. Such lasers have now  been routinely realized in experiments~\cite{Carder2003, Williams2006, Jin2020}.}

{We note that our formalism does not necessarily require a specific optical frequency, as only the detuning relative to the cavity frequency is relevant. 
However, a good choice of optical frequency is important in the design of the chiral cavity, as it needs to allow the necessary free spectral range of the cavity and keep in mind the associated limitations of cavity size and dimensions of the ferromagnetic strips inside the cavity.
%
% one chooses to work with, the corresponding FSR of the cavity will set some limitations to the size of the cavity, and thus of the ferromagnetic strip dimensions that can be placed in them.
% depending on the resonant optical frequency one chooses to work with, the corresponding FSR of the cavity will set some limitations to the size of the cavity, and thus of the ferromagnetic strip dimensions that can be placed in them.
%
Moreover, with lasers in the infrared to visible regime, the operation is possible with slightly larger power requirements.}

% \section{Optical power\label{app:power}}

%%%
% We plot this spectrum for different roots in Fig.~\ref{fig:spectral}.

\bibliography{Bibliography-CTC}

\end{document}